\documentclass[apj]{emulateapj}
\usepackage{array} 
\usepackage{amsmath}
\usepackage{wasysym}

\graphicspath{{Figs/}}

\newcommand\chandra{{\it Chandra}}
\newcommand\ciao{CIAO}
\newcommand\caldb{CALDB}

\newcommand\xspec{XSPEC}
\newcommand\eg{e.g.}
\newcommand\nelec{n_{\rm e}}
\newcommand\ntot{n_{\rm tot}}
\newcommand\emiss{\varepsilon}
\newcommand\rshock{r _{\rm s}}

\newcommand\nh{n_{\rm H}}
\newcommand\vinit{V_{\rm i}}
\newcommand\vfin{V_{\rm f}}
\newcommand\rin{r_{\rm i}}
\newcommand\rout{r_{\rm o}}
\newcommand\pshock{p_{\rm ps}}
\newcommand\prim{p_{\rm rim}}
\newcommand\residshock{\chi_{\rm ps}^2}
\newcommand\residrim{\chi_{\rm rim}^2}
\newcommand\dang{D_{\rm A}}
\newcommand\ie{i.e.}

\newcommand\phsp[1][]{p_{\rm hs#1}}
\newcommand{\phse}{\phsp[,E]}
\newcommand{\phsw}{\phsp[,W]}
\newcommand\phsmin[1][]{\phsp[,min#1]}
\newcommand\phsmine{\phsmin[,E]}
\newcommand\phsminw{\phsmin[,W]}
\newcommand\pjet{P_{\rm j}}
\newcommand\pjmax{P_{\rm j, max}}
\newcommand\rjet{r_{\rm j}}
\newcommand\kTjet{kT_{\rm j}}

\newcolumntype{T}{>{\raggedleft\arraybackslash}p{6em}}

\newenvironment{tightcenter}{%
  \setlength\topsep{0pt}
  \setlength\parskip{0pt}
  \begin{center}
}{%
  \end{center}
}

\shorttitle{Cocoon Shock of Cyg~A}
\shortauthors{Snios et al.}

\begin{document}

\title{The Cocoon Shocks of Cygnus A: Pressures and Their Implications
  for the Jets and Lobes}
	
\author{Bradford Snios$^{1}$}
\author{Paul E. J. Nulsen$^{1,2}$}
\author{Michael W. Wise$^{3,4}$}
\author{Martijn de Vries$^{4}$}
\author{Mark Birkinshaw$^{5}$}
\author{Diana M. Worrall$^{5}$}
\author{Ryan T. Duffy$^{5}$}
\author{Ralph P. Kraft$^{1}$}
\author{Brian R. McNamara$^{6}$}
\author{Chris Carilli$^{7}$}
\author{Judith H. Croston$^{8}$}
\author{Alastair C. Edge$^{9}$}
\author{Leith E. H. Godfrey$^{3}$}
\author{Martin J. Hardcastle$^{10}$}
\author{Daniel E. Harris$^{1,*}$}
\altaffiliation{${}^{*}$Dan Harris passed away on 2015 December 6. 
	His contributions to radio and X-ray astronomy will always be remembered.}
\author{Robert A. Laing$^{11}$}
\author{William G. Mathews$^{12}$}
\author{John P. McKean$^{3}$}
\author{Richard A. Perley$^{7}$}
\author{David A. Rafferty$^{13}$}
\author{Andrew J. Young$^{5}$}

\affil{${}^{1}$Harvard-Smithsonian Center for Astrophysics, 60 Garden Street,
	Cambridge, MA 02138 USA}
\affil{${}^{2}$ICRAR, University of Western Australia, 35 Stirling Hwy,
	Crawley, WA 6009, Australia}
\affil{${}^{3}$ASTRON (Netherlands Institute for Radio Astronomy), P.O. Box 2, 7990 
	AA Dwingeloo, The Netherlands}
\affil{${}^{4}$Astronomical Institute ``Anton Pannekoek," University of Amsterdam,
	Postbus 94249, 1090 GE Amsterdam, The Netherlands}
\affil{${}^{5}$H. H. Wills Physics Laboratory, University of Bristol, Tyndall Ave, 
	Bristol BS8 1TL, UK}
\affil{${}^{6}$Department of Physics \& Astronomy, University of Waterloo, 200 
	University Avenue, West Waterloo, Ontario N2L 3G1, Canada}
\affil{${}^{7}$National Radio Astronomy Observatory, P.O. Box 0, Socorro, 
	NM 87801, USA}
\affil{${}^{8}$School of Physical Sciences, The Open University, Walton Hall, 
	Milton Keynes, MK6 7AA, UK}
\affil{${}^{9}$Department of Physics, University of Durham, South Road, Durham, 
	DH1 3LE, UK}
\affil{${}^{10}$School of Physics, Astronomy and Mathematics, University of 
	Hertfordshire, College Lane, Hatfield, Hertfordshire AL10 9AB, UK}
\affil{${}^{11}$Square Kilometre Array Organisation, Jodrell Bank Observatory, 
	Lower Withington, Macclesfield, Cheshire SK11 9DL, UK}
\affil{${}^{12}$UCO/Lick Observatory, Department of Astronomy and Astrophysics, 
	University of California, Santa Cruz, CA 95064, USA}
\affil{${}^{13}$Hamburger Sternwarte, Universit{\"a}t Hamburg, Gojenbergsweg 112, 
	D-21029, Hamburg, Germany}

\begin{abstract}
We use 2.0 Msec of \chandra{} observations to investigate the cocoon 
shocks of Cygnus~A  and some implications for its lobes and jet. 
Measured shock Mach numbers vary in the range 1.18--1.66 around 
the cocoon. We estimate a total outburst energy  of 
$\simeq 4.7\times10^{60}\rm\ erg$, with an age of $\simeq 2 \times 10^{7}\rm\ yr$. 
The average postshock pressure is found to be 
$8.6 \pm 0.3 \times 10^{-10}\rm\ erg\ cm^{-3}$, which agrees with the 
average pressure of the thin rim of compressed  gas between the 
radio lobes and shocks, as determined from X-ray spectra. However, 
average rim pressures are found to be lower in the western lobe 
than in the eastern lobe by $\simeq 20\%$. Pressure estimates for hotspots A and D 
from synchrotron self-Compton models imply that each jet exerts a 
ram pressure $\gtrsim 3$ times its static pressure, consistent with 
the positions of the hotspots moving about on the cocoon shock over time. 
A steady, one-dimensional flow model is used to 
estimate jet properties, finding mildly relativistic flow 
speeds within the allowed parameter range. Models in which 
the jet carries a negligible flux of rest mass are
consistent with with the observed properties of the jets 
and hotspots. This favors the jets being light, implying that
the kinetic power and momentum flux are carried primarily by 
the internal energy of the jet plasma rather than by its rest mass.
\end{abstract}

\keywords{galaxies: active -- galaxies: clusters: general -- 
	galaxies: individual (Cygnus A) -- X-rays: galaxies}

%%%%%%%%%%%%%%%%%%%%%%%%%%%%%%%%%%%

\section{Introduction} 
\label{sec:intro}

It is widely accepted that active galactic nuclei (AGNs) significantly
affect their galaxy hosts, likely playing a central role in the
formation and evolution of galaxies and larger-scale structure
\citep[\eg,][]{Fabian2012}. For many galaxy clusters, such as those with 
cool cores, in the absence of a heat source, X-ray-emitting hot gas at the 
center would start cooling in less than 1 Gyr, at rates in excess of 
one hundred solar masses per year. However, in the majority of 
cases, a radio AGN hosted by the central galaxy deposits sufficient 
power via jets to prevent the gas from cooling \citep{Birzan2004, 
Dunn2006, Rafferty2006, McNamara2007}. By limiting copious 
cooling and the consequent star formation at cluster centers, radio 
AGNs can resolve the cooling flow problem \citep{Fabian1994, 
Tabor1993, Tucker1997}, account for the lack of star formation
in the central galaxies, and explain the steep decline in the galaxy
luminosity function at high luminosities \citep{Bower2006, Croton2006}.
Similar phenomena are observed in the lower mass halos of galaxy
groups and massive elliptical galaxies that host a substantial hot
atmosphere. As a result, the interaction between radio AGNs hosted by
cluster central galaxies and their environments has become a central
issue for structure formation.

The Fanaroff--Riley class II (FRII) radio galaxy \citep{Fanaroff1974}
Cygnus A (Cyg~A) is the archetype of powerful radio galaxies
\citep{Carilli1996}. At a redshift of $z = 0.0561$ 
\citep{Owen1997, Smith2002, Duffy2018} and 
with an estimated jet power approaching $10^{46}\rm\ erg\ s^{-1}$
\citep[\eg,][and see below]{Godfrey2013}, it is by far the nearest truly
powerful radio galaxy in the universe. Cyg~A is hosted by the central
galaxy of a rich, cool-core galaxy cluster \citep{Owen1997}, and X-ray
observations can provide a valuable probe of the energy flows through
the jets from its AGN, the interaction of the jets with the surrounding 
medium, and the overall system's impact on its environment 
\citep[\eg,][]{Carilli1988,Carilli1994, Harris1994, Smith2002, Rafferty2006}. 
X-ray observations of Cyg~A also provide a unique opportunity to investigate 
the physical structure of a powerful radio galaxy and discuss its evolution
over time. Beyond further understanding of Cyg~A, analysis of 
this system is also beneficial to the study of FRII systems in general.

This paper is one of a series on the analysis and interpretation of
2.0 Msec of \chandra{} X-ray observations of Cyg~A. Its focus is the cocoon
shocks of Cyg~A and what they tell us about the AGN outburst and the
physical properties of the lobes and jets. The cocoon shocks extend
from $\simeq 30\arcsec$ (33 kpc) north of the AGN, at their closest,
to just beyond the western hotspot, at $\simeq 67\arcsec$ (74 kpc)
from the AGN on the sky. They are driven by the momentum and power
deposited by the jets \citep{Scheuer1974, Begelman1984, Heinz1998, 
Reynolds2001}. 
In Section~\ref{sec:observation}, we describe the data used and
outline our method of data analysis. In
Section~\ref{sec:deprojection}, we determine the radial profiles of the
properties of the intracluster medium (ICM) in sectors centered on the
AGN. In Section~\ref{sec:shockfits}, we expand upon previous works
\citep[\eg][]{Smith2002, Wilson2006}, using X-ray surface brightness
profiles to determine shock strengths at a number of locations around
the periphery of the cocoon. The results rely on model-dependent
assumptions, which we test by checking consistency with several
alternative measures of the shock strength, based on temperature
jumps, shock compression, and postshock pressures. Postshock pressures
provide good estimates of the pressure within the radio lobes, away
from the immediate vicinity of the hotspots, where the pressure is
expected to be substantially higher than in the rest of the lobe
\citep{Scheuer1974, Harris1994, Carilli1996, Blundell1999,
  Mathews2012}. Postshock pressures are presented in
Section~\ref{sec:directpres}, together with pressure estimates 
determined from X-ray spectra for the narrow rim of compressed 
gas between the cocoon shock and the radio lobes. 
Some physical consequences of our results are discussed in
Section~\ref{sec:discussion}. Results for the rate of expansion of
the cocoon shock and a self-similar model for the inflation of the
radio lobes are used to estimate the velocity of the AGN relative to
the hot gas and the speed of advance of the hotspots. The fitted
shock models are used to estimate the outburst energy and mean power
of the jets. Lastly, estimates of the hotspot pressures from
synchrotron self-Compton models are used to obtain estimates of 
the jet speeds.

We assume $H_{0} = 69.3\rm\ km\ s^{-1}\ Mpc^{-1}$, $\Omega_{M} = 0.288$, 
and $\Omega_{\Lambda} = 0.712$ \citep{Hinshaw2013}, which give an angular
scale for Cyg~A of $1.103\rm\ kpc\ arcsec^{-1}$ and an angular
diameter distance of  227 Mpc at the redshift $z = 0.0561$. The Galactic 
H\,I column density is set to $3.1 \times 10^{21}\rm\  cm^{-2}$ based on 
an average of the results from \citet{Dickey1990} and \citet{Kalberla2005}. All 
uncertainty ranges are 68\% confidence intervals, unless otherwise stated. 

%%%%%%%%%%%%%%%%%%%%%%%%%%%%%%%%%%%

\section{\textit{Chandra} Data Reduction} 
\label{sec:observation}

\begin{figure*}
	\begin{tightcenter}
	\includegraphics[width=0.96\textwidth]{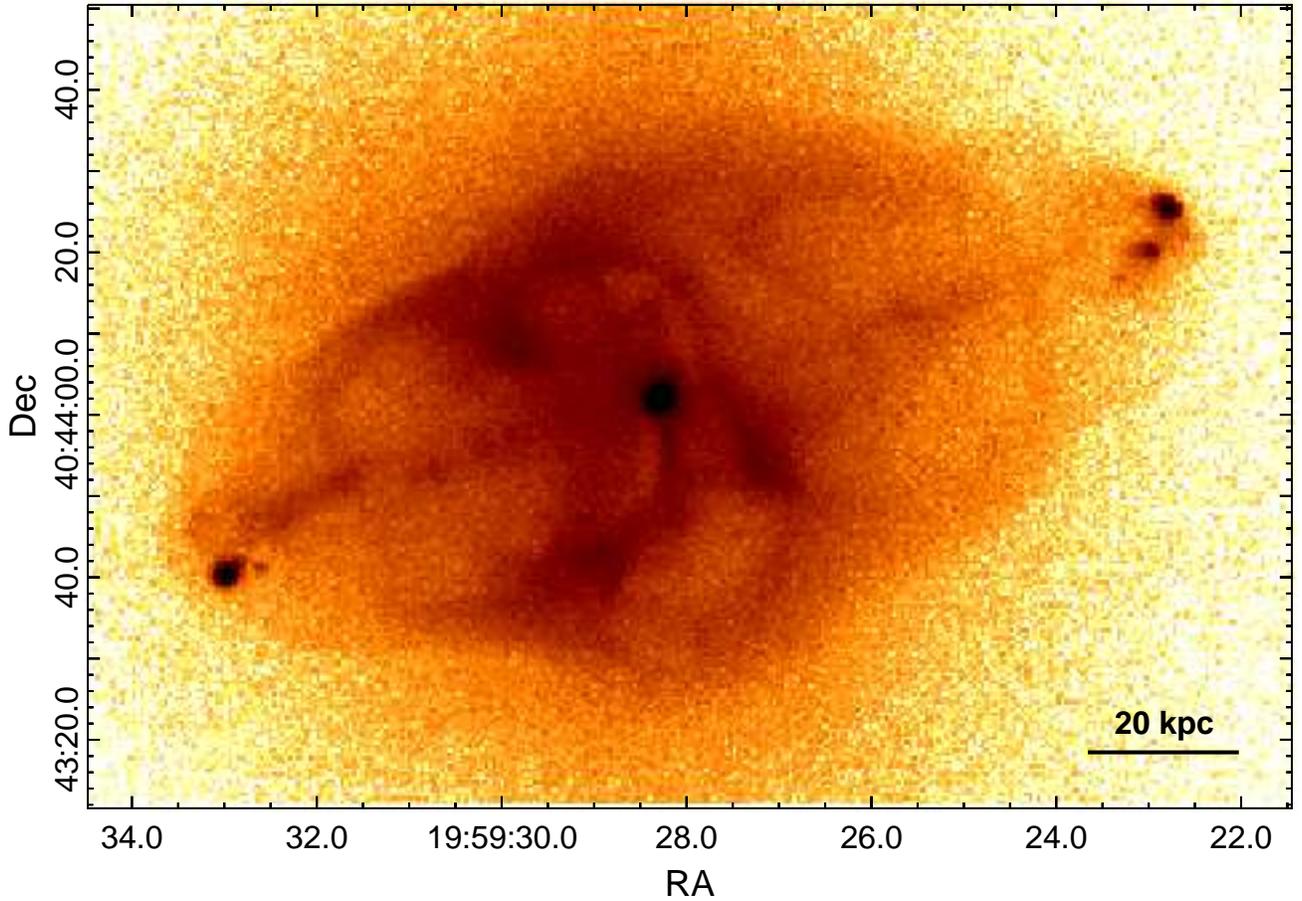}
	\caption{0.5 -- 7.0 keV \chandra{} image of Cygnus A. The 
	image has been background-subtracted and exposure-corrected 
	and is made using the \chandra{} observations listed in Table \ref{table:obs}.}
  	\label{fig:cyga}
	\end{tightcenter}
\end{figure*}

\begin{figure}
	\begin{tightcenter}
	\includegraphics[width=0.46\textwidth]{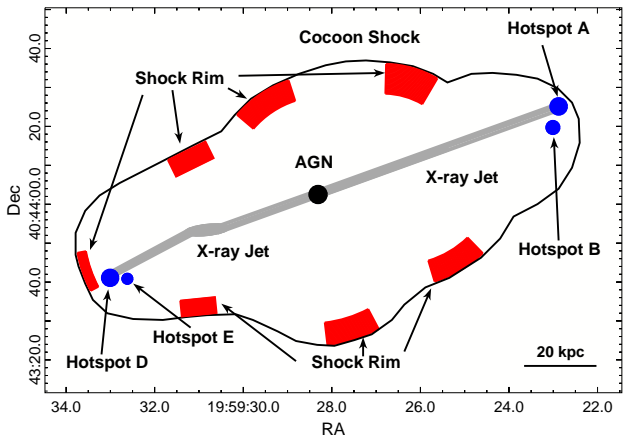}
	\caption{Schematic image of Cygnus A with key features (AGN, 
		cocoon shock, hotspots, jets) labeled for ease of visibility. 
		Several inner rims of the shock front that are visible in 
		the original image are also highlighted.}
  	\label{fig:cyga2}
	\end{tightcenter}
\end{figure}

\begin{table}
	\caption{\textit{Chandra} Observations Used}
	\label{table:obs}
	\begin{tightcenter}
	{\footnotesize 
		\begin{tabular}{ c c c | c c c }
		\hline \hline
		ObsID & Date & $T_{\rm exp}$\tablenote{Net exposure after background flare removal.} & ObsID & Date & $T_{\rm exp}{}^{a}$ \\
		 &  & (ks) &  &  & (ks) \\
		\hline
		00360 & 2000-05-21 & 34.3 & 17517 & 2016-09-17 & 26.7 \\
		01707 & 2000-05-26 & 9.2 & 17518 & 2016-07-16 & 49.4 \\
		05830 & 2005-02-22 & 23.5 & 17519 & 2016-12-19 & 29.6 \\
		05831 & 2005-02-16 & 50.6 & 17520 & 2016-12-06 & 26.8 \\
		06225 & 2005-02-15 & 24.3 & 17521 & 2016-07-20 & 24.7 \\
		06226 & 2005-02-19 & 23.6 & 17522 & 2017-04-08 & 49.4 \\
		06228 & 2005-02-25 & 15.8 & 17523 & 2016-08-31 & 49.4 \\
		06229 & 2005-02-23 & 22.6 & 17524 & 2015-09-08 & 22.8 \\
		06250 & 2005-02-21 & 7.0 & 17525 & 2017-04-22 & 24.5 \\
		06252 & 2005-09-07 & 29.7 & 17526 & 2015-09-20 & 49.4 \\
		17133 & 2016-06-18 & 30.2 & 17527 & 2015-10-11 & 26.3 \\
		17134 & 2017-05-20 & 28.5 & 17528 & 2015-08-30 & 49.1 \\
		17135 & 2017-01-20 & 19.8 & 17529 & 2016-12-15 & 34.9 \\
		17136 & 2017-01-26 & 22.2 & 17530 & 2015-04-19 & 21.1 \\
		17137 & 2017-03-29 & 25.0 & 17650 & 2015-04-22 & 28.2 \\
		17138 & 2016-07-25 & 26.0 & 17710 & 2015-08-07 & 19.8 \\
		17139 & 2016-09-16 & 39.5 & 18441 & 2015-09-14 & 24.6 \\
		17140 & 2016-10-02 & 34.2 & 18641 & 2015-10-15 & 22.4 \\
		17141 & 2015-08-01 & 29.7 & 18682 & 2015-10-14 & 22.6 \\
		17142 & 2017-04-20 & 23.3 & 18683 & 2015-10-18 & 15.6 \\
		17143 & 2015-09-03 & 26.9 & 18688 & 2015-11-01 & 34.4 \\
		17144 & 2015-05-03 & 49.4 & 18871 & 2016-06-13 & 21.6 \\
		17507 & 2016-11-12 & 32.6 & 18886 & 2016-07-23 & 22.2 \\
		17508 & 2015-10-28 & 14.9 & 19888 & 2016-10-01 & 19.5 \\
		17509 & 2016-07-10 & 51.4 & 19956 & 2016-12-10 & 54.3 \\
		17510 & 2016-06-26 & 37.1 & 19989 & 2017-02-12 & 41.5 \\
		17511 & 2017-05-10 & 15.9 & 19996 & 2017-01-28 & 28.1 \\
		17512 & 2016-09-15 & 66.9 & 20043 & 2017-03-25 & 29.6 \\
		17513 & 2016-08-15 & 49.4 & 20044 & 2017-03-26 & 14.9 \\
		17514 & 2016-12-13 & 49.4 & 20048 & 2017-05-19 & 22.6 \\
		17515 & 2017-03-21 & 39.3 & 20077 & 2017-05-13 & 27.7 \\
		17516 & 2016-08-18 & 49.0 & 20079 & 2017-05-21 & 23.8 \\
		\hline
		\multicolumn{5}{r}{Total Exposure Time} & 1958.7 \\
		\hline
	\end{tabular}}
	\end{tightcenter}
\end{table}

Cyg~A was initially observed by \chandra{} on 2000 May 21 (ObsID 00360)
using the Advanced CCD Imaging Spectrometer (ACIS) with the object
centered on the S3 chip in FAINT mode. A follow-up observation was
performed with the S3 chip in VFAINT mode (ObsID 01707), and all
subsequent observations were performed with ACIS-I centered on the AGN
of Cyg~A, its western hotspot, or its eastern hotspot (observations
targeting the merging subcluster were not used). A complete list
of the observations used in our analysis is given in
Table~\ref{table:obs}. All data were reprocessed using \ciao{} 4.9,
with \caldb{} 4.7.4 \citep{Fruscione2006}, and the routine deflare was used to
remove background flares. The resulting cleaned exposure times are
shown in Table \ref{table:obs}, with a total exposure time of 1.96
Msec. Additionally, the readout\_bkg routine was used to
estimate the distribution of ``out-of-time'' events, those due to
events detected during frame transfer, for each observation. The
cleaned exposures, corrected for out-of-time events, were used in all
of the analysis discussed in this article.

In order to correct for small astrometric errors, ObsID 05831 was
chosen as a reference for its high total count. A raw $\mbox{0.5 -- 
7.0} \rm\ keV$ image was made in a rectangular region of $160\arcsec \times
120\arcsec$ centered on Cyg~A. For each remaining ObsID, the events were
reprojected onto the sky to match ObsID 05831, and a raw \mbox{0.5 -- 
7.0} keV image was made for the same region. The cross-correlation
between each raw image and ObsID 05831 image was then fitted with a
Lorentzian profile to determine the offset between them. The 
astrometric translation required to align each data set with ObsID 05831 
was then applied to the event list using the wcs\_update \ciao{} routine. 
The root mean square translation for the images was 
$\Delta {\rm x}_{\rm rms} = 0.82\arcsec$ 
and $\Delta{\rm y}_{\rm rms} = 0.25\arcsec$. This approach produced notably 
sharper features in a co-added image of Cyg~A than those using the 
\ciao{} tools designed to coalign the point sources.

The appropriate blank-sky exposures from CALDB were processed in
an analogous fashion to the data to simulate a
background event file for each observation. The background rates
were scaled to match observed rates in the \mbox{10 -- 12} keV energy
band. Exposure maps for the \mbox{0.5 -- 7.0} keV energy band were
created assuming the spectral model \textsc{phabs} $\times$
\textsc{apec}, with a temperature $kT = 5.5\rm\  keV$ and an abundance
\mbox{$Z = 0.66$} relative to the solar abundances of
\citet{Anders1989}. A \mbox{0.5 -- 7.0} keV, background-subtracted,
exposure-corrected image made from the combined exposures is shown in
Figure~\ref{fig:cyga}. The cocoon shock is clearly seen enveloping the
eastern and western hotspots and the other complex structure that
surrounds the central AGN (Figure~\ref{fig:cyga2}).
  
All spectra used in the following analysis were binned to have a
minimum of 1 count per bin and are fitted over the energy range
\mbox{0.5 -- 7.0} keV using the Cash statistic (cstat) in \xspec{}
v12.9.1h \citep{Arnaud1996}. Abundances are scaled to the solar 
abundances of \citet{Anders1989}.
 
%%%%%%%%%%%%%%%%%%%%%%%%%%%%%%%%%%%

\section{Properties of the Unshocked ICM} 
\label{sec:deprojection}

\begin{figure}
	\begin{tightcenter}
	\includegraphics[width=0.46\textwidth]{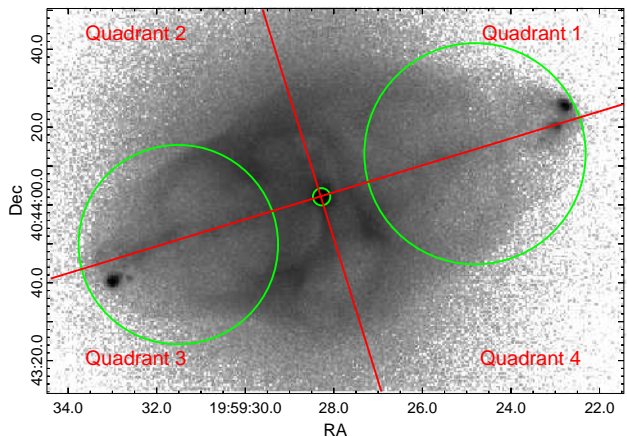}
	\end{tightcenter}
        \caption{Regions used for deprojections. The red lines show
          the boundaries of the four quadrants. Green circles show
          the regions excluded. Data for each quadrant were analyzed
          separately, and quadrants 2-4 were also deprojected together.
          Quadrant 1 is excluded from the latter group, since it is
          most affected by the infalling subcluster.}
	\label{fig:quad}
\end{figure}

\begin{figure*}
	\begin{tightcenter}
	\includegraphics[width=0.50\textwidth]{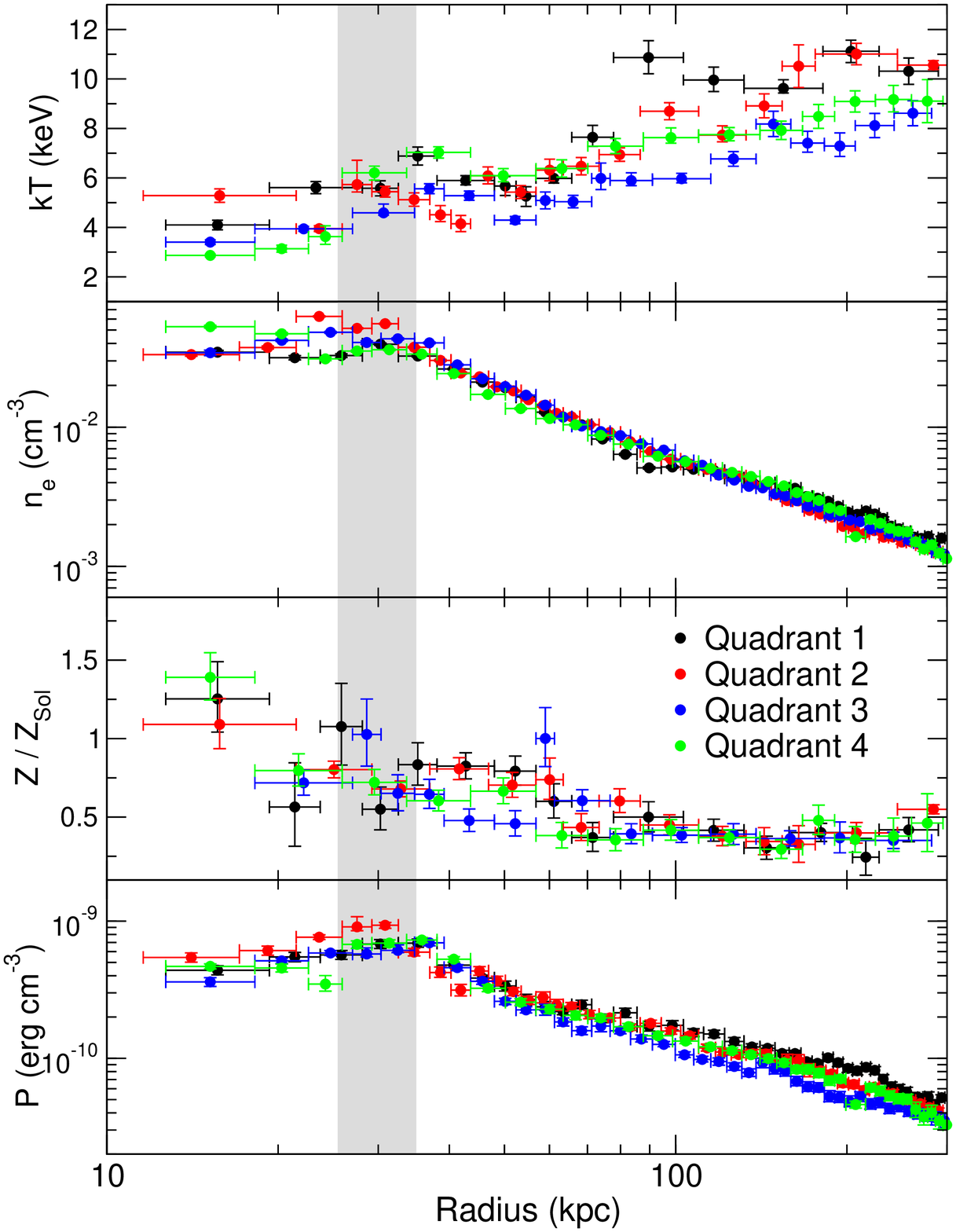}\includegraphics[width=0.50\textwidth]{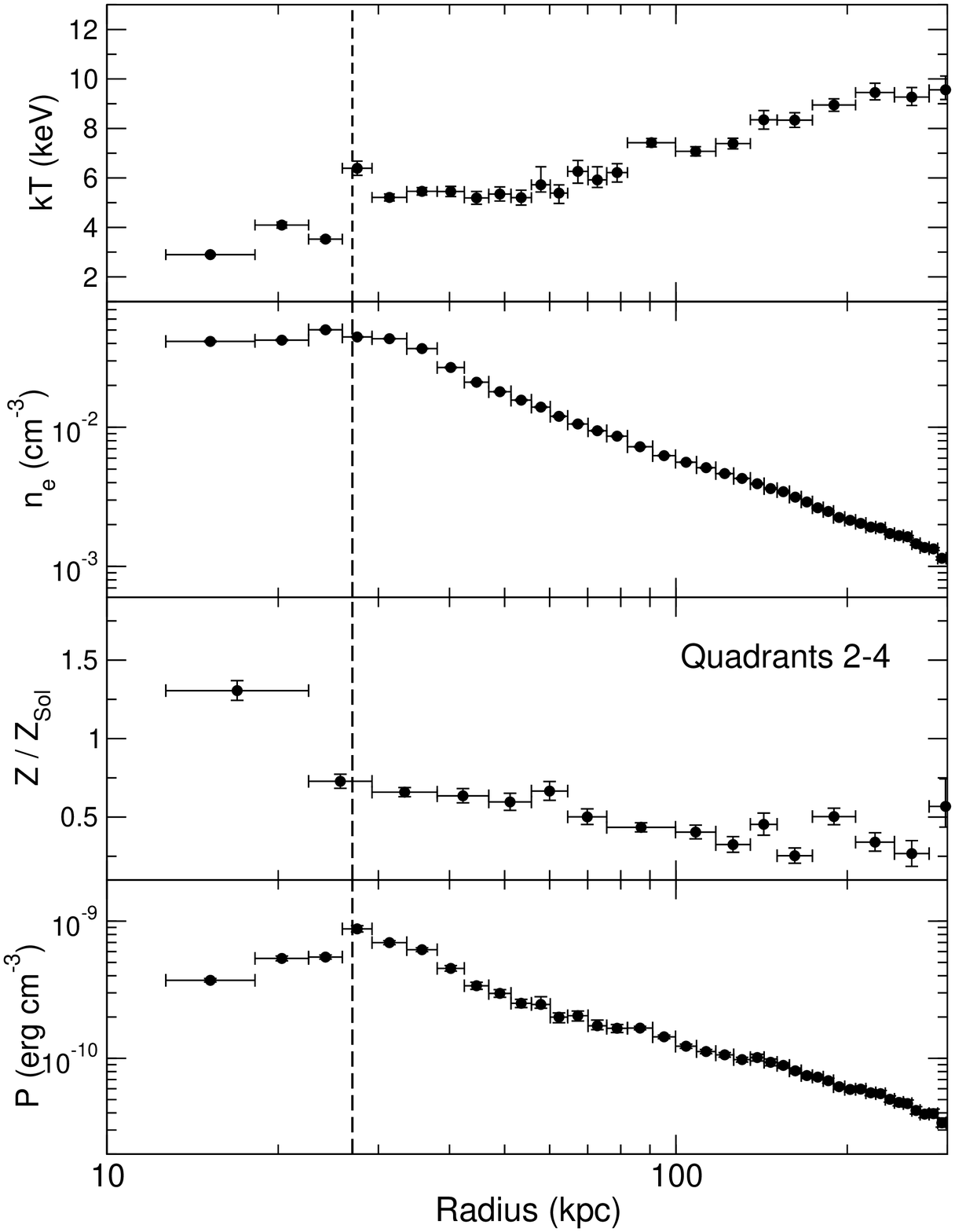}
	\end{tightcenter}
	\caption{Deprojected profiles of the temperature,
                electron density, abundance, and pressure for the four
                quadrants marked in Figure~\ref{fig:quad}
                (left) and for quadrants 2 -- 4 combined
                (right). The gray stripe indicates the range
                of shock radii, while the dashed line shows the
                average shock radius.}
	\label{fig:deprojection}
\end{figure*}

Deprojections were used to determine the properties of the unshocked
ICM in the vicinity of the cocoon shock. The region around Cyg~A was
divided, along the jets and a perpendicular axis, into quadrants about
the AGN, as shown in Figure~\ref{fig:quad}. A region around the central
AGN and circular regions over the lobes and hotspots were masked.
Annular sectors were then defined in each quadrant, out to a radius of 
$300\arcsec$, to have a minimum of 25,000 counts in each. Annular sectors 
were also defined to cover quadrants 2 -- 4, containing a minimum of 75,000
counts per region. The northwest sector (quadrant 1) was excluded, as
it is most affected by the merger shock associated with the infalling
subcluster \citep{Ledlow2005, Wise2018}

For the deprojections, each set of annular spectra was fitted
simultaneously using the \xspec{} model $\textsc{projct} \times
\textsc{phabs} \times \textsc{mekal}$. Two additional thermal
components are included as a second model. The first of these allows
for emission from the cluster atmosphere beyond the deprojection
region, under the assumption that the gas there is distributed as an
isothermal beta model \citep{Nulsen2010}. The beta parameter for this
model was determined by fitting the surface brightness profile from
$200\arcsec$ to $300\arcsec$ in each quadrant. The second component of the
additional model represents soft thermal emission of uniform surface
brightness, to allow for foreground emission from our Galaxy. The
deprojection provides temperatures and abundances directly, while
values of the electron densities, $\nelec$, are determined from the
norms of the thermal models determined with \textsc{projct}, assuming that
the density is uniform in the spherical shells. Total gas pressures
are determined as $\ntot kT$, where the total particle number
density is $\ntot \simeq 1.93 \nelec$.

The deprojection results are plotted in Figure~\ref{fig:deprojection}, and the 
corresponding data tables are provided in the supplemental materials. 
Average values are in excellent agreement with the deprojection results  
of \cite{Smith2002}. The radius of the cocoon shock varies between 
and within the sectors, in the range indicated by the gray band in
Figure~\ref{fig:deprojection}. Within the shock, the deprojection
results must be treated with caution due to the evident departures
from spherical symmetry, which is discussed in more detail 
by \cite{Duffy2018}. Beyond it, the electron density profiles are
remarkably consistent from quadrant to quadrant, only showing modest
departures from spherical symmetry. The largest discrepancy occurs at
$\simeq 80\arcsec$ in the northwest quadrant (quadrant 1), where the density
is roughly 30\% lower than in the other quadrants and the temperature
is markedly higher. There is more scatter between the temperatures in
the different quadrants, which are significantly higher at larger
radii in the northern sectors and lowest to the southeast, on the
opposite side to the infalling subcluster. At a radius of $200\arcsec$, the
pressures span a range of $\simeq 2$ from northwest to southeast,
which reduces substantially at smaller radii, outside the shock. The
temperature profiles each show a modest local peak at about the shock
radius. The abundance errors largely obscure any structure, apart
from an overall decline with radius.

%%%%%%%%%%%%%%%%%%%%%%%%%%%%%%%%%%%

\section{Cocoon Shock Strength}
\label{sec:shockfits}

In this section, we determine the shock strength at several locations
on the cocoon shock by fitting its surface brightness profile.
Temperature jumps and shock compression are considered for consistency
checks. Later, in Section \ref{sec:directpres}, we also examine the
pressure jump in each sector, comparing it to a direct estimate of the
gas pressure within the cocoon.

\begin{figure*}
	\begin{tightcenter}
	\includegraphics[width=0.51\textwidth]{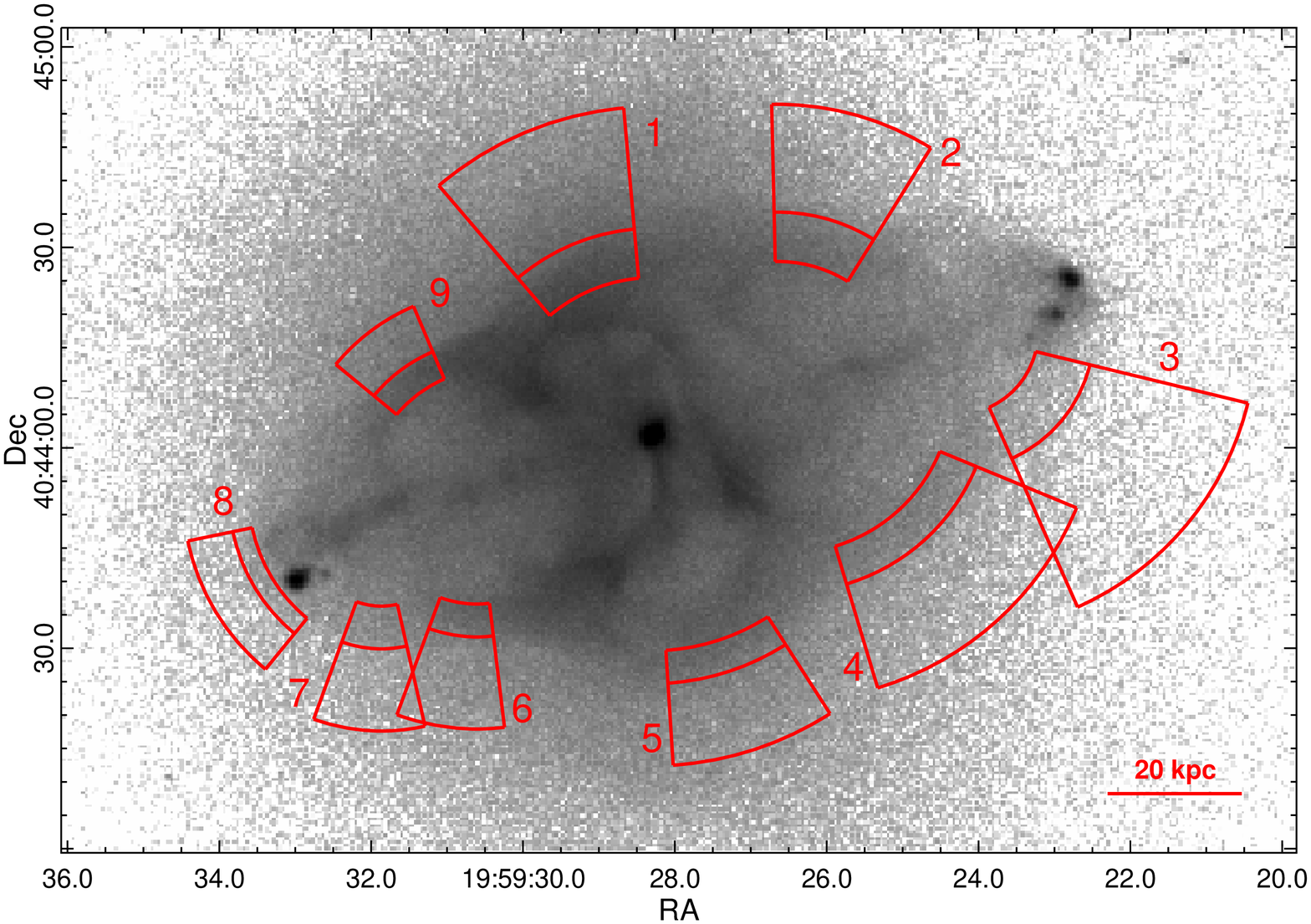}\includegraphics[width=0.46\textwidth]{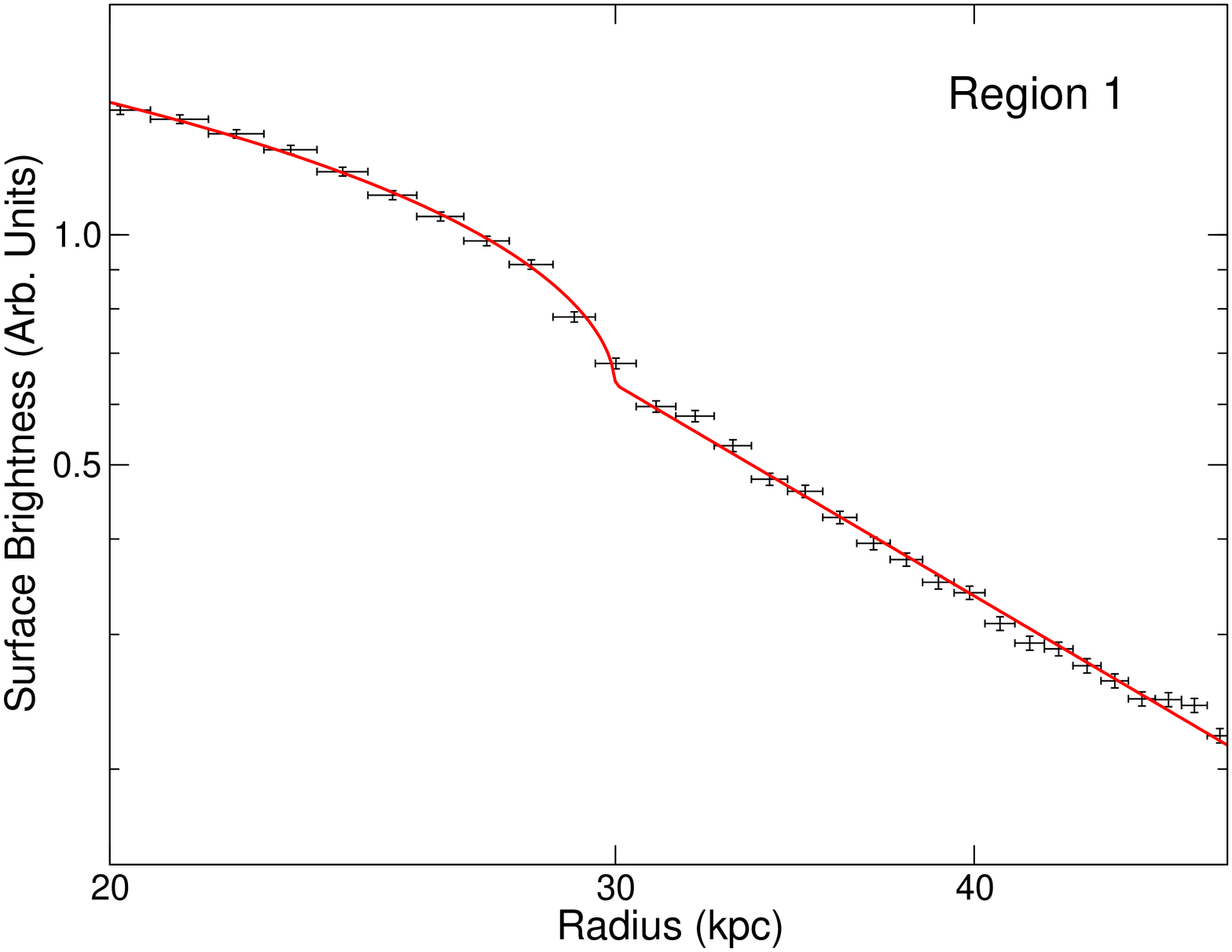} 
	\end{tightcenter} 		
        \caption{Left: sectors used to measure the surface
          brightness profiles of the cocoon shock. Inner and outer
          arcs mark the range of the radius fitted, and the middle arc
          marks the best-fit radius of the density discontinuity.
          Right: Surface brightness
          profile and best-fit hydrodynamic model for sector 1
          (Section \ref{sec:hydroshock}).}
	\label{fig:regions}
\end{figure*}

To measure the surface brightness profiles of the cocoon shock, a number
of segments of the shock front were chosen where the shock is clearly
visible and continuous. A sector was defined to enclose each segment 
such that an arc in the sector best matches the front. This procedure gave 
the nine sectors marked in Figure~\ref{fig:regions}. A surface brightness
profile of the shock was extracted for each sector. To model the
surface brightness profile of a shock front, its radius of curvature
relative to our line of sight is critical, as that determines how
much of the line of sight lies within the shocked gas at any projected
distance from the front. In practice, the radius of curvature is
determined by the coordinate used in the surface brightness profile.
The zero point of this coordinate therefore needs to be chosen
suitably. Under the assumption that the cocoon shock is axially
symmetric about the X-ray jet, the center of the jet is 
chosen as the center of every sector. The units of the radial 
coordinate do not affect the estimated shock strength, so there
is no need to correct for the inclination between each sector and
the axis of the cocoon.

The models for the surface brightness profile assume that the emission
arises from hot gas. All profiles were therefore truncated prior to entering 
cavities and/or regions of nonthermal emission within
the shock. At larger radii, the profiles were truncated before any
noticeable departure from a constant power-law slope. Each 
sector was also selected to avoid evident structure, such as 
the region directly to the south of the AGN where the gas rim 
appears narrowest. Limiting the transverse extent of a sector
reduces the signal-to-noise ratio of the surface brightness profile,
while limiting the radial extent of the fit generally increases the
uncertainty in the fitted parameters. Nevertheless, the parameters
are well-determined in all the regions selected. The limits of
each sector are shown, together with an arc marking the fitted 
shock, in Figure~\ref{fig:regions}.

%%%%%%%%%%%%%%%%%%%%%%%%%%%%%%%%%%%

\subsection{Broken Power-law Fits to Surface Brightness Profiles}
\label{sec:broken}

The broken power-law model for the surface brightness profile is obtained 
by assuming that the distribution of volume emissivity is spherical in
three dimensions, with the form
\begin{equation}
  \emiss(r) =
  \begin{cases}
    A_1 (r /\rshock)^{-\eta_1},& \text{for $r < \rshock$}, \\
    A_2 (r / \rshock)^{-\eta_2},& \text{for $r > \rshock$},
  \end{cases}
\end{equation}
and the constant parameters $A_1$, $A_2$, $\eta_1$, $\eta_2$ and
$\rshock$, where $\rshock$ is the shock radius. Projecting the volume
emissivity onto the sky (by integrating along the line of sight) gives the
model surface brightness profile, which is binned and fitted to the observed
profile to determine the parameters. For thermal plasma, the power
radiated per unit volume is $\nelec \nh \Lambda (T, Z)$, where the
cooling function $\Lambda$ depends on the temperature $T$ and
composition $Z$ of the gas. The composition is expected to vary
slowly near the shock and the \chandra{} broadband response for
thermal plasma depends very weakly on the temperature in the range of
interest, so the brightness is almost independent of the
temperature. Since the proton number density, $\nh$, is a constant
multiple of the electron number density, we can therefore estimate
the density jump at the shock as $\sqrt{A_1/A_2}$. The shock Mach
number is then determined from the density jump using the
Rankine-Hugoniot jump conditions for gas with the ratio of specific
heats, $\Gamma = 5/3$. The density jumps and Mach numbers 
obtained from the broken power-law model are given for the nine
sectors in the columns 3 and 4 of Table~\ref{table:sbp}.

\begin{table*}
	\caption{Shock Parameters}
	\label{table:sbp}
	\begin{tabular}{ c | c | c c | c c | c c c c }
		\hline \hline
		 & & \multicolumn{2}{c|}{Broken Power Law} & \multicolumn{2}{c|}{1D Hydro Model} \\
		\hline
		Region\tablenote{Region number from Figure~\ref{fig:regions}.}  & Discontinuity & Density & Mach & Density & Mach & Abundance\tablenote{Relative to the scale of \citet{Anders1989}.}
  & $kT_{\rm out}$ & $kT_{\rm in}$ & $kT_{\rm in} / kT_{\rm out}$ \\
		& Distance\tablenote{Average projected distance from the AGN to the shock front.}
 (kpc) & Jump & Number & Jump &
			Number & & (keV) & (keV) & \\ 
		\hline
		1 & $32.9 \pm 0.2$ & $1.31\substack{+0.02\\-0.02}$ & $1.21\substack{+0.01\\-0.01}$
			& $1.27\substack{+0.01\\-0.01}$ & $1.18\substack{+0.01\\-0.01}$ & 
			$0.60\substack{+0.03\\-0.03}$ & $5.16\substack{+0.06\\-0.06}$ & 
    			$5.26\substack{+0.05\\-0.05}$ & $1.02\substack{+0.02\\-0.02}$ \\ 
		2 & $46.1 \pm 0.7$ & $1.52\substack{+0.03\\-0.01}$ & $1.35\substack{+0.01\\-0.02}$ 
			& $1.58\substack{+0.03\\-0.01}$ & $1.40\substack{+0.02\\-0.01}$ & 
			$0.65\substack{+0.06\\-0.06}$ & $7.87\substack{+0.24\\-0.24}$ & 
    			$7.47\substack{+0.22\\-0.21}$ & $0.95\substack{+0.06\\-0.05}$ \\
		3 & $68.3 \pm 0.4$ & $1.55\substack{+0.11\\-0.07}$ & $1.38\substack{+0.08\\-0.05}$ 
			& $1.79\substack{+0.12\\-0.03}$ & $1.56\substack{+0.10\\-0.03}$ & 
			$0.52\substack{+0.08\\-0.07}$ & $9.09\substack{+0.44\\-0.44}$ & 
    			$8.95\substack{+0.49\\-0.40}$ & $0.98\substack{+0.11\\-0.08}$ \\
		4 & $48.7 \pm 0.4$ & $1.56\substack{+0.04\\-0.04}$ & $1.38\substack{+0.04\\-0.03}$ 
			& $1.66\substack{+0.03\\-0.03}$ & $1.46\substack{+0.02\\-0.02}$ & 
			$0.44\substack{+0.04\\-0.04}$ & $7.54\substack{+0.18\\-0.18}$ & 
    			$8.08\substack{+0.23\\-0.22}$ & $1.07\substack{+0.06\\-0.05}$ \\ 
		5 & $41.0 \pm 0.3$ & $1.31\substack{+0.04\\-0.05}$ & $1.21\substack{+0.03\\-0.03}$ 
			& $1.43\substack{+0.03\\-0.03}$ & $1.29\substack{+0.02\\-0.02}$ & 
			$0.44\substack{+0.03\\-0.04}$ & $5.74\substack{+0.13\\-0.13}$ & 
    			$6.05\substack{+0.17\\-0.16}$ & $1.05\substack{+0.04\\-0.05}$ \\
		6 & $45.6 \pm 0.3$ & $1.58\substack{+0.06\\-0.05}$ & $1.40\substack{+0.04\\-0.03}$ 
			& $1.67\substack{+0.04\\-0.04}$ & $1.47\substack{+0.03\\-0.03}$ & 
			$0.68\substack{+0.05\\-0.05}$ & $5.31\substack{+0.10\\-0.10}$ & 
    			$6.98\substack{+0.28\\-0.25}$ & $1.31\substack{+0.09\\-0.07}$ \\
		7 & $57.3 \pm 0.2$ & $1.72\substack{+0.22\\-0.15}$ & $1.51\substack{+0.15\\-0.11}$ 
			& $1.82\substack{+0.09\\-0.08}$ & $1.58\substack{+0.07\\-0.06}$ & 
			$0.61\substack{+0.06\\-0.06}$ & $5.82\substack{+0.18\\-0.18}$ & 
    			$7.66\substack{+0.41\\-0.40}$ & $1.32\substack{+0.11\\-0.11}$ \\
		8 & $70.7 \pm 0.5$ & $1.82\substack{+0.38\\-0.23}$ & $1.58\substack{+0.33\\-0.18}$ 
			& $1.87\substack{+0.14\\-0.17}$ & $1.62\substack{+0.12\\-0.13}$ & 
			$0.49\substack{+0.08\\-0.07}$ & $6.29\substack{+0.24\\-0.23}$ & 
			$7.05\substack{+0.43\\-0.35}$ & $1.12\substack{+0.11\\-0.09}$ \\
		9 & $43.0 \pm 0.2$ & $1.90\substack{+0.10\\-0.09}$ & $1.65\substack{+0.06\\-0.06}$ 
			& $1.92\substack{+0.07\\-0.07}$ & $1.66\substack{+0.06\\-0.06}$ & 
			$0.65\substack{+0.06\\-0.06}$ & $5.51\substack{+0.17\\-0.14}$ & 
    			$6.99\substack{+0.30\\-0.20}$ & $1.27\substack{+0.09\\-0.07}$ \\  
		\hline
	\end{tabular}
\end{table*}

%%%%%%%%%%%%%%%%%%%%%%%%%%%%%%%%%%%

 \subsection{Hydrodynamic Shock Model}
\label{sec:hydroshock}

The surface brightness profiles were also fitted using the spherical
hydrodynamic model described in \citet{Nulsen2005}. In this model,
the unshocked gas is assumed to be isothermal and hydrostatic, with a
power-law density distribution, $\rho(r) \propto r^{-\eta}$. A shock
is launched by an initial, explosive energy release at the center of
the grid and the subsequent gas flow is calculated using a spherically
symmetric hydrodynamic code. The preshock temperature is chosen to
match the temperature of the gas just outside the shock. Note
that, for this model, the \chandra{} response is included in computing
the surface brightness profiles to fit to the data. Since the model is
scale-free, it can be rescaled at each time step to obtain the best fit
to the surface brightness profile. Optimizing the fit over time for
one simulation gives a best-fitting shock radius and Mach number.
Simulations are then run for a range of initial density power laws,
$\eta$, to find the global best fit.

Although this model still represents a highly simplified version of the
cocoon shock in Cyg~A, it provides a better account than the broken
power-law model of the rapid expansion of the shocked gas that occurs
immediately after the shock. As a more physically accurate model, we
therefore prefer its results to those for the broken power-law model.
However, the broken power-law model has been used widely, so it is
interesting to compare the results. Density jumps and Mach numbers
for the hydrodynamic model are given in columns 5 and 6 of
Table~\ref{table:sbp} and an example fit for region 1 is shown in
Figure~\ref{fig:regions}. Although the differences are marginal in most
cases, the shock strengths for the hydrodynamic model are
systematically higher than those for the broken power-law model, except 
for region 1. Given the greater fidelity of the hydrodynamic model, the 
results indicate that the broken power-law fits tend to systematically
underestimate the shock strength, although by a small amount for these
relatively weak shocks. Results from the hydrodynamic model are used in 
the remainder of this article. 

%%%%%%%%%%%%%%%%%%%%%%%%%%%%%%%%%%%

\subsection{Shock Temperature Jumps} 
\label{sec:shockfronttemp} 

Two spectra were extracted from each of the nine sectors shown in
Figure~\ref{fig:regions}, from the regions inside and outside the fitted
shock radius. Temperatures were determined by 
fitting the pre- and postshock spectra with the single-temperature
model $\textsc{phabs} \times \textsc{apec}$ in \xspec{}. Abundances
were assumed to be the same on either side of the shock in each
sector, as we do not expect large local variations. Temperatures and
normalizations were allowed to vary independently. Since emission from 
the unshocked gas is projected onto the region inside the shock,
a two-temperature model was also tried for the region within the
shock, with the temperature of the one thermal component tied to that of
the region outside the shock. However, this model did not
significantly improve the fits. The abundances and fitted
temperatures for each sector are given in Table~\ref{table:sbp}, with
values from within the shock denoted $kT_{\rm in}$ and those from
outside denoted $kT_{\rm out}$.

For $\Gamma = 5/3$, the temperature jump in a weak shock is
numerically close to the value of its Mach number (for $M=1.18$, the
temperature jump is $1.17$, while for $M=1.66$, it is $1.67$), so
the Mach numbers from Table~\ref{table:sbp} should be directly
comparable to the temperature ratios in its last column.
However, several factors reduce the jumps in the projected temperature. 
First, unshocked gas projected onto the postshock region is generally 
cooler than the shocked gas, which lowers the fitted postshock 
temperatures. Second, adiabatic expansion causes a rapid
decrease in the gas temperature behind the shock, so that the finite
width of the postshock spectral regions inevitably makes their mean
temperatures lower than the immediate postshock temperatures. Third,
from the deprojected temperature profiles (Figure~\ref{fig:deprojection}), 
the shock appears to be propagating up a preexisting temperature 
gradient, which would now make the gas in the preshock region now hotter 
than the gas in the postshock region was before being shocked. All 
three effects tend to make the jump measured in the projected 
temperature lower than the jump at the shock front. Thus, the measured 
temperature ratios should be regarded as lower limits on the actual shock 
temperature jumps. Although the measured jumps do not provide good 
quantitative measures of the shock strength, taken together, they make 
a strong case that the temperature increases in the shock. Furthermore, 
for the Mach numbers determined from the surface brightness profiles, 
the measured ratios in the projected temperature are broadly consistent 
with the expectations of numerical models \citep[\eg,][]{Forman2007}. 
In particular, the temperature jumps are higher in the sectors with higher 
Mach numbers.

\begin{figure}
	\begin{tightcenter}
	\includegraphics[width=0.46\textwidth]{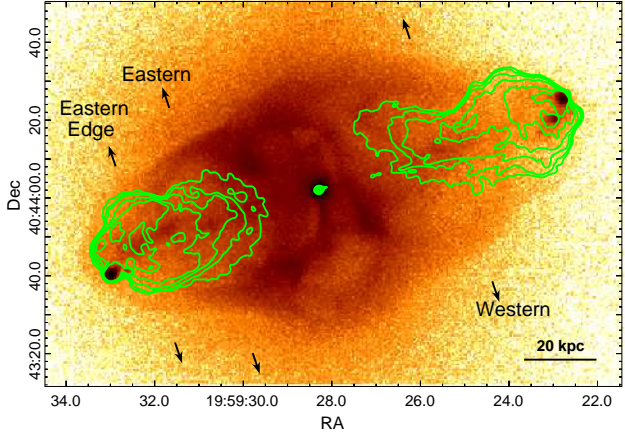}
	\end{tightcenter}
	\caption{0.5 -- 7.0 keV \chandra{} image of Cygnus A
  		with 5 GHz VLA (green) contours overlaid. The contours start 
		at $3\sigma$ and are spaced by a factor of two. Radio emission fills 
		the eastern and western cavities, and the central radio peak is 
		coincident with the AGN. Arrows indicate the positions of the surface 
		brightness cuts through the cocoon discussed in Sections 
		\ref{sec:compression} and \ref{sec:nonthermal}.}
	\label{fig:radio}
\end{figure}

%%%%%%%%%%%%%%%%%%%%%%%%%%%%%%%%%%%

\subsection{Shock Compression}
\label{sec:compression}

\begin{figure}
	\begin{tightcenter}
	\includegraphics[width=0.47\textwidth]{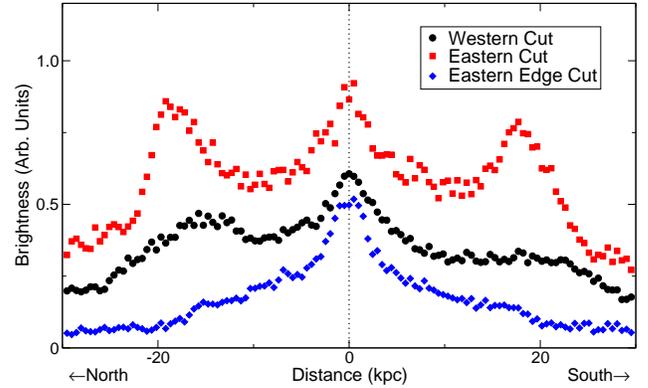}
	\end{tightcenter}
	\caption{Surface brightness cuts perpendicular to the radio axis.
		The positions of the three cuts are indicated by arrows in
		Figure~\ref{fig:regions}. Each cut through the radio cocoon is 
		centered on the peak over the X-ray jet. The bright rims of
		compressed gas bounding the X-ray ``cavity'' on the north and
		south are clearest for the eastern cut (red). Further to the east
		(blue), there appear to be no bright rims, and the surface
		brightness rises monotonically toward the center of the jet,
		requiring diffuse X-ray emission from throughout the lobe. To the
		west, the northern rim is less distinct and no rim is evident to
		the south.}
	\label{fig:cut}
\end{figure}

Within the cocoon shock, there is a clear anticorrelation between the
X-ray emission and the 5 GHz radio emission, as shown in Figure~\ref{fig:radio}. 
This and the detailed matches between radio and X-ray features
around the edges of the lobes make a strong case that 
the radio plasma has displaced the X-ray-emitting gas, 
as found in many less powerful radio galaxies \citep[\eg,][]{McNamara2007, 
Wise2018}. The compression of the gas into narrow rims
around the radio lobes is clearest in the eastern lobe, between
regions 6 and 9 of Figure~\ref{fig:regions}. This is also seen in the
surface brightness cut through the cocoon in this region, plotted in
red in Figure~\ref{fig:cut}.

Here we ask whether the compression is consistent with the estimated
shock strengths for regions 6 and 9. To estimate how much the
displaced gas has been compressed, we first assume that all of the gas
initially within the lobe remains in the rims (rather than being
displaced toward the cluster center, for example). We also assume
cylindrical symmetry about the axis of the X-ray jet. For a fixed
amount of gas, the mean density is inversely proportional to the
volume, so that the compression is given by $\vinit / \vfin$, where
$\vinit$ and $\vfin$ are the initial and final volumes occupied by the
gas. If the gas was displaced perpendicular to the jet axis, we would
have $\vinit / \vfin = \rout^2 / (\rout^2 - \rin^2)$ (cylindrical
motion), where $\rin$ and $\rout$ are the inner and outer radii of the
compressed shell. More generally, as the gas is pushed away from the jet
axis, fluid elements will also separate in the direction along the
axis. If the separation increases linearly with distance from the
axis, the volume would scale as $r^2 + \beta r^3$, for some constant
$\beta \ge 0$. In the limit $\beta r \gg 1$, we would then have
$\vinit / \vfin = \rout^3 / (\rout^3 - \rin^3)$, corresponding to
spherical motion (note that $\beta r \ll 1$ gives the cylindrical case).

For region 6, we estimate that the perpendicular distance from the jet axis
to the inner edge of the compressed rim is $\rin = 17\farcs6$ and to
the shock front it is $\rout = 24\farcs6$, giving compressions ranging
from 1.58 to 2.05 for the spherical and cylindrical cases,
respectively. From the Rankine-Hugoniot jump conditions, for Mach
number of 1.47, the shock compression would be a factor of 1.67, which lies in
this range. For region 9, we find $\rin = 16\farcs4$ and $\rout = 22\farcs1$, 
giving compressions in the range 1.69 -- 2.23. For a Mach number of 
1.66, the shock compression is a factor of 1.92, which is also within
the estimated range. We note that these results are rough, and other
issues, such as the rapid expansion of the gas behind the shock and likely
variation of the shock strength over time, add systematic uncertainty. 
Despite these concerns, it is reassuring that the Mach numbers 
determined from the surface brightness profiles are consistent with 
our estimates of the compression of the gas in the rims.

%%%%%%%%%%%%%%%%%%%%%%%%%%%%%%%%%%%

\subsection{Diffuse Lobe Emission}
\label{sec:nonthermal}

Consider a cylindrical shell of uniform X-ray emission, between the inner 
and outer radii $a$ and $b$, respectively. Projecting onto the sky (in
any direction but parallel to the axis of the cylinder), the surface
brightness on a line of sight that passes within a distance $\varpi < a$
of its axis will be proportional to $\sqrt{b^2 - \varpi^2} - \sqrt{a^2
  - \varpi^2}$, which is an increasing function of $\varpi$.
Decomposing any cylindrically symmetric distribution of X-ray emission
into thin cylindrical shells of uniform emission, this shows that if
there is a hollow central region, the surface brightness will always
increase with distance away from the symmetry axis inside the hollow
region. Although the issue is more complex for a more general
axisymmetric distribution of X-ray emission, it generally remains true
that the surface brightness must increase with distance from the
symmetry axis if the central region is hollow.

This is the basis of the discussion in Section \ref{sec:compression}
above. Particularly in the inner eastern part of the cocoon, we see
clear evidence that the radio plasma has displaced the hot ICM,
compressing it into a dense shell between the radio lobe and the
cocoon shock. This is manifested in the surface brightness cut (red
points in Figure~\ref{fig:cut}) as sharp decreases in the surface
brightness inside the northern and southern rims. Emission from the
central ``X-ray jet'' partly fills in the X-ray cavity, but the
decrease in X-ray surface brightness within the rims shows that any
X-ray emission from inside the cavity must be considerably fainter
than that from within the rims.

Farther toward the eastern hotspots, although the shock front remains
quite visible in the X-ray image, the compressed shell of shocked gas
is no longer readily discernible. This is confirmed by the ``eastern
edge'' surface brightness cut in Figure~\ref{fig:cut} (blue points),
which shows that the X-ray emission rises inwards, all the way to the
central peak over the X-ray jet. Such a surface brightness profile is
inconsistent with a hollow shell of X-ray emission. It requires that
there is diffuse X-ray emission throughout the lobe region, peaking
toward the jet axis. Almost certainly, the centrally peaked X-ray
emission from within the bright radio lobe is the synchrotron
self-Comptom emission reported previously 
\citep{Hardcastle2010, Yaji2010, deVries2018}. Given that
the shock is likely to be stronger here than in regions closer to the AGN 
(due to higher lobe pressure in the vicinity of the hotspots and to 
lower external pressure), we should expect the shell of the 
shock-compressed gas to be thinner relative to the size of
the cavity than it is in regions closer to the AGN. There must be
some thermal emission from the shocked gas, but it is hard to
distinguish from nonthermal emission from within the lobe. As a
result, it is difficult 
to know where the thermal shock model is applicable and, therefore, to
determine the shock strength. For this reason, the regions used to
measure the shock profiles (Figure~\ref{fig:regions}) lie outside the radio
lobes, although this was not possible for region 3.

%%%%%%%%%%%%%%%%%%%%%%%%%%%%%%%%%%%

\section{Cocoon Pressure}
\label{sec:directpres}

In this section, we consider estimates of the pressure within the
radio cocoon, which is a key parameter of physical models
for the lobes. Jet momentum can maintain higher pressures in the
hotspots, driving supersonic flows and nonuniform pressures in small
surrounding regions \citep[\eg,][]{Mathews2012}. However, the 
sound speed is expected to be very high in the plasma filling the
radio lobes, so that the pressure should be relatively uniform away
from the vicinity of the hotspots. The pressure is also expected to
be fairly uniform within the rim of compressed gas between the radio
lobes and the cocoon shock, so that the pressure in the rim can
provide a good measure of the pressure in the adjacent lobe.

\begin{table}[t]
	\caption{A Comparison of Shock Pressures}
	\label{table:pressure}
	\begin{tightcenter}
	\begin{tabular}{ c | T l | T l }
		\hline \hline
		& \multicolumn{4}{c}{Pressure ($10^{-10}$ erg cm$^{-3}$)} \\
		\hline
		Region &\multicolumn{2}{c|}{Postshock\tablenote{Determined from preshock and shock strength, 
  		Section \ref{sec:pspressure}.}} &
			\multicolumn{2}{c}{Rim\tablenote{Determined from \xspec{} norms of spectra, 
  		Section \ref{sec:rimpressure}.}} \\
		\hline
		1 & $9.59\substack{+0.83\\-0.80}$ & & $8.44\substack{+0.21\\-0.21}$ & \\
      		2 & $8.46\substack{+0.69\\-0.66}$ & & $8.34\substack{+0.39\\-0.38}$ & \\	
		3 & $5.74\substack{+0.93\\-0.85}$ & & $7.12\substack{+0.60\\-0.55}$ & \\
		4 & $7.83\substack{+0.70\\-0.67}$ & & $8.48\substack{+0.58\\-0.40}$ & \\
		5 & $9.02\substack{+0.75\\-0.72}$ & & $9.07\substack{+0.45\\-0.43}$ & \\
		6 & $8.93\substack{+0.88\\-0.83}$ & & $10.77\substack{+0.73\\-0.65}$ & \\
		7 & $6.51\substack{+1.19\\-1.08}$ & & $9.59\substack{+0.50\\-0.49}$ & \\
		8 & $5.57\substack{+1.37\\-1.18}$ & & $6.17\substack{+0.65\\-0.58}$ & \\
		9 & $10.04\substack{+1.88\\-1.69}$ & & $12.16\substack{+0.91\\-0.73}$ & \\
      		\hline
	\end{tabular}
	\end{tightcenter}
\end{table}

%%%%%%%%%%%%%%%%%%%%%%%%%%%%%%%%%%%

\subsection{Postshock Pressure}
\label{sec:pspressure}

Our first determination of the pressures in the lobes relies on the
shock fits. For each of the regions marked in Figure~\ref{fig:regions},
the deprojected pressure profiles of Section \ref{sec:deprojection}
can be used to estimate a preshock pressure. The deprojected distance
from a shock to the AGN is determined from the projected shock radius
in Table~\ref{table:sbp}, assuming that the cocoon axis is inclined at
$55^\circ$ to the line of sight \citep{Vestergaard1993}. Thus, the
displacement from the AGN parallel to the jet axis is boosted by 
a factor of $1/\sin 55^\circ$, while the displacement perpendicular 
to the axis is unaltered. The preshock pressure is taken from the 
deprojected pressure profile for the appropriate quadrant at the 
deprojected distance from the AGN. For the sectors that cross 
between two quadrants (regions 1, 5, and 8), the pressure values 
are averaged for those quadrants. The Mach number for     
the hydrodynamic model (Table~\ref{table:sbp}) is then used to
calculate the pressure jump in the shock, hence the postshock
pressure. The resulting postshock pressures are listed in the
second column of Table~\ref{table:pressure}, and their range is modest. 
If they are consistent with a single value, $\pshock$, the residual
\begin{equation} \label{eqn:residual}
\residshock = \sum_i \left(\frac{p_i - \pshock}{\sigma_i}\right)^2
\end{equation}
should have an approximately chi squared distribution. It is
minimized by setting $\pshock$ to the weighted mean, 
\begin{equation} \label{eqn:wmean}
\pshock =  \frac{\sum_i p_i / \sigma_i^2}{\sum_i 1 / \sigma_i^2}.
\end{equation}
Using this value in the residual reduces the number of degrees of
freedom by one. Excluding the outlying values for regions 3 and 8  and
taking  
$\sigma_i$ to be the average of the upper and lower sigmas for each 
measurement, the weighted mean postshock pressure is 
$\pshock = 8.56 \pm 0.31 \times10^{-10}\rm\ erg\ cm^{-3}$,
giving the residual $\residshock = 7.28$, near the
70\% upper confidence limit for a chi-squared distribution with six
degrees of freedom. Thus, the postshock pressures do not show
evidence for pressure variations within the cocoon

Several sources of systematic error inherent to our model
affect the postshock pressures. First, the geometry of the 
system is fixed by assuming axial symmetry about 
the center of the X-ray jet. This fixes the shape of the front
and the spatial distribution of the gas, which determine a surface
brightness profile. While theory and the appearance of Cyg~A support
the assumption of axial symmetry, it is clearly approximate. In
particular, local irregularities on the front can be caused by
gas flows within the ICM, stochastic precession of the jet, or
variations in jet power \citep{Heinz2006, Mendygral2012}. 
Small-scale irregularities have the effect of smoothing the projected 
surface brightness profile, making the shock appear weaker 
\citep{Nulsen2013}. Larger-scale asymmetries can alter the 
curvature of the projected front. The impact of such effects is 
expected to be greater on the outer parts of the front (regions 3, 7 and 8), 
where the scale of the intrinsic curvature is smaller, causing the front 
seen in projection to appear less clearly defined. They may well 
account for the anomalous postshock pressures of region 3 and 8. 

A second related source of systematic error is the implicit assumption 
that the densest unshocked gas on our line of sight coincides with the
projected shock front. This maximizes the density of the gas
being shocked, hence the contrast in surface brightness at the shock.
If it is incorrect, our surface brightness fits underestimate
the true shock strength. The asymmetry of the X-ray image and the
evidence that the AGN is moving through the ICM (Section
\ref{sec:motion}) both make it unlikely that this assumption is
completely accurate. 

A third source of systematic error for the postshock pressures 
is the poorly constrained inclination angle of the jet axis. Various inclination 
angles have been used in prior Cyg~A analyses, ranging from 35$^\circ$--80$^\circ$ 
\citep{Bartel1995, Boccardi2016}. Recalculating the postshock pressures 
using the minimum and maximum angles of this range produced a $\sim$20\% 
decrease and increase in the average postshock pressure, respectively.
The estimates of rim pressures in the next section rely on fewer assumptions, 
providing some check on these sources of systematic error.

%%%%%%%%%%%%%%%%%%%%%%%%%%%%%%%%%%%

\subsection{Rim Pressure}
\label{sec:rimpressure}

The rims of the eastern cavity, in particular, are significantly
brighter than the adjacent unshocked gas. This suggests that the 
temperature and density in a rim can be estimated by simply by ignoring 
the emission of the unshocked gas projected onto the rim. Doing so
overestimates the gas density in the rim, hence also its pressure. A
more accurate result might be obtained by deprojection, where we
estimate how much emission from adjacent regions is
projected onto the rim and, in effect, treat it as a background
contribution to the rim spectrum. However, the results of
deprojection are sensitive to unavoidable assumptions about the
distribution of the unshocked gas. Rather than attempting to model
the distribution of unshocked gas, we can simply treat the adjacent
region as a local background, almost certainly
overestimating the amount of emission projected onto the rim. The
upshot will be to underestimate the pressure in the rim. Combining
these two estimates provides lower and upper bounds on the
pressure in the rim, which can bracket its pressure tightly when the
rim is much brighter than the adjacent, unshocked region.

We apply this approach to the spectra that were used to determine the 
pre- and postshock temperatures in Section \ref{sec:shockfronttemp}. Each
sector in Figure~\ref{fig:regions} is divided at the shock radius, and
two spectra are extracted. The spectrum of the inner region
represents emission from the compressed rim, and it is used to
determine gas properties in the rim by fitting the absorbed thermal
model, $\textsc{phabs} \times \textsc{apec}$, in \xspec. Using a blank-sky 
background, the fit gives us an upper limit on the pressure, while
using the preshock spectrum for background gives a lower limit.
Treating the gas in the rim as uniform, its density can be determined
in cgs units from the \xspec{} norm,
\begin{equation}
\textsc{norm} = \frac{10^{-14} \nelec \nh V}{4 \pi [\dang (1 + z)]^2},
\end{equation}
where $\dang$ is the angular diameter distance, $z$ is redshift 
and $V$ is the volume of the emitting region, and the proton number
density is $\nh = 0.86\nelec$. To estimate the emitting volume, we
assume again that the rim is symmetric under rotation about the jet axis.
For a spectrum extracted from an annular sector, the emitting volume
then lies in the intersection between a spherical shell and a
cylindrical shell extending to infinity along our line of sight, with
the same inner and outer radii, restricted to the angular range of
the sector, $\delta\phi$. The volume of the region is therefore
\begin{equation}
V = \frac{2 \delta\phi}{3} (\rout^2 - \rin^2)^{3/2},
\end{equation}
where $\rin$ and $\rout$ are the inner and outer radii of the 
rim. As in Section \ref{sec:deprojection}, the rim pressure is $\ntot kT$,
with $\ntot = 1.93\nelec$, and the temperature determined from the
spectral fit. The right-hand column of Table~\ref{table:pressure}
gives the average of the upper and lower limits on the pressures, with
systematic errors equal to half the difference between the limits
combined in quadrature with the lower and upper confidence ranges.

Omitting the outlying value for region 8, the weighted mean of the rim
pressures is (Equation~\ref{eqn:wmean}) $\prim = 8.72 \pm 0.14 \times
10^{-10}\rm\ erg\ cm^{-3}$, in agreement with the mean postshock
pressure (Section \ref{sec:pspressure}). However, using this value to
compute the residual (Equation~\ref{eqn:residual}) gives $\residrim =
40.88$ for seven degrees of freedom, exceeding the 99\% confidence level
and indicating that the pressure does vary significantly around the
rim. From Table~\ref{table:pressure}, the pressures are higher in the
eastern lobe. The weighted mean of the rim pressures for the eastern regions 
(6, 7 and 9) is $10.41 \pm 0.36 \times10^{-10}\rm\ erg\ cm^{-3}$, while that
for the central and western regions (1, 2, 3, 4 and 5) is $8.41 \pm
0.15 \times 10^{-10}\rm\ erg\ cm^{-3}$, differing by $5\sigma$. 
The corresponding weighted mean of the postshock pressures for
the east is (regions 6, 7, and 9) $8.31 \pm 0.64 \times
10^{-10}\rm\ erg\ cm^{-3}$ and, for the center and west (regions 1, 2,
4 and 5) is $8.64 \pm 0.36 \times 10^{-10}\rm\ erg\ cm^{-3}$,
respectively. Thus, the mean postshock pressure for the central and
western regions is within $1 \sigma$ of that for the eastern regions,
whereas the mean rim pressure for the eastern lobe is almost
$3 \sigma$ higher than the mean postshock pressure.

%%%%%%%%%%%%%%%%%%%%%%%%%%%%%%%%%%%

\section{Discussion} 
\label{sec:discussion}

\subsection{Cocoon and Lobe Pressures}
\label{sec:uniformity}

As discussed in Section \ref{sec:pspressure}, the postshock pressures
are consistent with the single value of $8.6 \pm 0.3 \times 10^{-10}\rm\
erg\ cm^{-3}$. This value agrees well with the weighted mean of the
independent rim pressures, $8.7 \pm 0.2 \times 10^{-10}\rm\ erg\
cm^{-3}$ (Section \ref{sec:rimpressure}). These values are also
consistent with the weighted mean of the rim pressures for the central
and western parts of the cocoon, $8.4\pm0.2\times10^{-10}\rm\ erg\
cm^{-3}$, but about 20\% lower than the weighted mean of the rim
pressures for the eastern region of the cocoon,
$10.4\pm0.4\times10^{-10}\rm\ erg\ cm^{-3}$. In most shock models, the
shocked gas undergoes rapid adiabatic expansion immediately behind the
shock, so that, if anything, we should expect the rim pressures to be
lower than the postshock pressures, suggesting that our estimates of
the postshock pressures may be low (as discussed in Section
\ref{sec:rimpressure}). However, the relative narrowness of the rim of
shock-compressed gas in Cyg~A indicates that the rim gas is moving at
a substantial fraction of the shock speed and so does not expand much
behind the shock (see Section \ref{sec:compression}).

The rim pressures of Section~\ref{sec:rimpressure} should provide the
best measure of the pressure within the radio lobes of Cyg~A. The
rims lie immediately adjacent to the lobes, where the sound speed is
expected to be high, so they should have very similar pressures.
These pressure estimates rely almost solely on the assumption that the
cocoon is axially symmetric. Although this is unlikely to be exact,
the density estimates scale as the reciprocal of the square root of
the depth of the emitting regions, making them insensitive to modest
departures from the assumed geometry. If, for example, the high
surface brightness of the rims in regions 6 and 9
(Figure~\ref{fig:regions}) was due to the lobe cross-section being
elliptical rather than circular, the ellipse would need to have an
axial ratio of $\simeq1.53$. Although this cannot be ruled out, it is
implausible. The brightness of the rims in regions 6 and 9 shows 
that the density there is almost certainly higher than in the other parts
of the rim, and the 20\% difference between the pressure of the western
lobe and the rest of the cocoon is unlikely to be due to systematic
error. 

Such a large pressure difference is difficult to
explain. The results of Section \ref{sec:jets} support the widely
held assumption that the sound speed in the lobes is much greater than
the sound speed in the ICM, hence the speed of the cocoon shock. This
should mean that the pressure within the lobes is nearly uniform away
from the hotspots. Although the merger shock is overrunning the
cluster core, it is also slow compared to 
the sound speed in the lobes and should have very little impact on
the pressure gradients within the lobe. 

In regions away from the hotspots, flow speeds within the lobes are
generally low compared to the sound speed of the radio plasma. 
Thus, we expect the plasma pressure to
be relatively uniform away from the immediate vicinity of the hotspots 
\citep{Mathews2010,Chon2012}. Our pressure measurements confirm this
expectation, at least to the level of $\simeq 20\%$. We do not see
evidence for higher pressures in the outer parts of the cocoon,
close to the hotspots. However, our pressure measurements are sparse
and less accurate in these regions (where densities are lower and the
radius of curvature of the cocoon is smaller, reducing the brightness
contrast of the shock fronts).

Although the shock in region 8 is projected only 7.6 kpc beyond the
eastern hotspots, it cannot be associated directly with a hotspot. The
pressure in the radio lobes is expected to be highest in the hotspots,
so we expect the scale of the associated shock to be comparable to the
small size of the hotspot. As a result, thermal emission from a
hotspot shock will be very difficult to separate from its strong
nonthermal X-ray emission. The extent of the shock in region 8 is too
large for it to be part of the terminal jet shock. Its speed is also
too slow to be directly associated with the hotspot
(Table~\ref{table:sbp} and Section~\ref{sec:hotspotspeed}). This raises the
issue of how the shock in region 8 can be projected beyond the
hotspot, when its speed is significantly slower than the rate of
advance of the hotspot. The most likely explanation is that the shock
in region 8 is a transient feature. During most of its history, a
hotspot would have led the expansion to the east, as it does now to
the west. However, as they shift around in three dimensions, at times the
hotspots can be projected behind the projected leading edge of the
shock (see Section~\ref{sec:hotspotspeed}).

%%%%%%%%%%%%%%%%%%%%%%%%%%%%%%%%%%%
 
\subsection{Motion of the AGN Relative to the Gas}
\label{sec:motion}

Brightest cluster galaxies (BCGs) typically move at speeds exceeding 
$100\rm\ km\ s^{-1}$ with respect to their cluster hosts \citep{Lauer2014}.
Continuing merger activity also disturbs the hot ICM, setting it in
motion with respect to the cluster potential at speeds comparable to
the BCG or greater \citep{Ascasibar2006, Randall2011}. 
From Table~\ref{table:sbp}, the distance from the AGN to the
near part of the shock front to the north of the AGN in Cyg~A is 
$32.9 \rm\ kpc$, while the distance to the shock front to the south is 
$41.0 \rm\ kpc$. Combining the Mach numbers from Table \ref{table:sbp} 
with the deprojected preshock temperatures, the speeds of the shocks in 
regions 1 and 5 are $1500\rm\ km\ s^{-1}$ and $1670\rm\
km\ s^{-1}$, respectively. The observed difference in shock 
strengths may be due to the higher ICM density and pressure to the 
north, as shown in Figure~\ref{fig:deprojection}. Assuming 
that the average speed of separation of the shock fronts is constant at 
the current rate of $3170\rm\ km\ s^{-1}$, it would have taken 
$\simeq 2.28 \times10^{7}\rm\ yr$ for the shocks to reach their 
current separation. In that time, the southern front has moved 
$8.1\rm\ kpc$ farther from the AGN than the northern one, 
with a mean speed $348\rm\ km\ s^{-1}$ faster than the northern 
shock. At the outset, when the shock fronts were close together, we 
assume that the state of the ICM outside the shock front was the same 
to the north and south, so that the two shocks had the same speeds. 
If the shock speeds varied linearly with time, then the average difference 
in the shock speeds would have been $85\rm\ km\ s^{-1}$.
Attributing the remainder of the north--south asymmetry to the motion
northward, perpendicular to the cocoon axis, of the AGN, its northward
velocity would equal half of the remaining difference in the speeds,
\ie, $130\rm\ km\ s^{-1}$. 

The projected distances to the eastern shock and the western shock 
are $63.1\rm\ kpc$ and $74.3\rm\ kpc$, respectively, so that the 
western shock is $11.2\rm\ kpc$ farther from the AGN than the 
eastern one. For the age estimate above, this gives a mean speed 
difference along the cocoon axis projected onto the plane of the 
sky of $480\rm\ km\ s^{-1}$. As we lack an estimate for the difference 
in the shock speeds in this direction, we attribute the entire 
difference to the motion of the AGN through the gas, giving an
eastward velocity along the axis of $240\rm\ km\ s^{-1}$.

In the absence of a detailed model for the expansion history of the
lobes and hotspots, there is substantial systematic uncertainty in
both components of the estimated velocity. In particular, asymmetries
in the ICM pressure distribution can affect the shock speed. We therefore 
estimate a total AGN speed of $\simeq270\rm\ km\ s^{-1}$ with respect to the
gas, with a total systematic uncertainty of a factor of $\sim2$. 
Our projected speed is consistent with
the proper motion estimates from \cite{Steenbrugge2014}.

%%%%%%%%%%%%%%%%%%%%%%%%%%%%%%%%%%%

\subsection{Outburst Energy and Power}
\label{sec:ageenergy}

Scaling the spherical hydrodynamic shock model to match the data
enables all properties of the model to be expressed in physical units.
Doing this for the northern and southern shocks in the central region
(regions 1 and 5) provides two estimates for the age and energy of the
outburst driving the cocoon shock. These regions were used as
uncertainty in the geometry of the shock front is minimized where it is 
almost spherical. The temperature and density of the unshocked gas 
at radii of 40 kpc for the north and 45 kpc for the south were obtained 
from the deprojected profiles. For the northern shock, the age of the
outburst was found  
to be $t_{\rm N} = 1.87 \times 10^{7}\rm\ yr$, with a total energy  of
$E_{\rm N} = 6.67 \times 10^{59}\rm\ erg$. For the southern shock, the
age was $t_{\rm S} = 1.84 \times 10^{7}\rm\ yr$ and the total energy
$E_{\rm S} = 1.68 \times 10^{60}\rm\  erg$, a factor of $\sim2.5$ higher 
than for the north. Averaging the results produces a mean outburst 
age for the cocoon shock of $t_{\rm avg} = 1.86 \times 10^{7}\rm\ yr$, 
with a total energy of $E_{\rm avg} = 1.17 \times 10^{60}\rm\ erg$.

Most of the difference between the two energy estimates is due to the
difference in the volumes of the northern and southern shocked
regions. From Table~\ref{table:sbp}, the southern shock radius is
$\simeq 25\%$ larger than the northern shock radius. In the
spherically symmetric hydrodynamic model, this makes the volume
enclosed by the southern shock almost twice that enclosed
by the northern shock. All other things being equal, it would mean
that the southern shock requires twice as much energy. As argued in
Section \ref{sec:motion}, the AGN is moving northward through
the ICM, exaggerating the apparent difference in the shock radii.
Assuming that the outburst originates midway between the two shocks,
rather than at the current location of the AGN, reduces the difference 
in outburst energy to $\simeq 30\%$, with a comparable average energy 
to the original value. Much of the remaining energy difference can be 
attributed to the greater strength of the southern shock.

Additional uncertainty is present in the energy calculations because 
the volume of the spherical central region significantly underestimates 
the total volume of the shocked cocoon. A sphere of diameter 
equal to the distance from the northern shock to the southern shock 
has a volume of $\simeq 2.08 \times 10^{5} \rm\ kpc^{3}$. Assuming 
axial symmetry, we have estimated the volume of the shocked cocoon 
by measuring its width perpendicular to the cocoon axis
at many positions along the axis. Treating the cocoon as a stack of
sections of cones, its volume can be approximated as the sum of the
section volumes. Assuming that the cocoon axis is inclined at
$55^\circ$ to our line of sight, we correct for projection by 
boosting the result by a factor of $1/\sin 55^\circ$ to obtain 
a total volume of $4.08\times10^5\rm\ kpc^3$. Given that
the pressure within the lobes is approximately uniform (Sections
\ref{sec:pspressure} and \ref{sec:rimpressure}), the total shock
energy will have been underestimated by a factor close to the ratio of
this volume to that of the spherical central region, or $\simeq 2$.

A further shortcoming of our hydrodynamic model is that the outburst
is assumed to inject all of its energy explosively in a single,
initial event. This is unrealistic. As discussed in
\citet{Forman2017}, the history of energy release determines what
fraction of the energy resides in the lobes. This is minimized in a
single explosive outburst, which is clearly ruled out for the
lobes and cavities of Cyg~A. If the energy were deposited at a
constant rate instead, approximately twice as much energy would be
required to obtain the same shock strength \citep{Hardcastle2013,
  English2016}. In the absence of a more 
detailed model, we assume that the outburst power has been roughly
constant, so that the total energy estimate needs to be boosted by a
further factor of $\simeq2$ over the value from the explosive
hydrodynamic model. Putting the corrections together, we
expect that we have underestimated the total outburst energy by a
factor of $\simeq 4$. The systematic error in this is unlikely to
exceed a factor of 2. Thus, we estimate the total outburst energy
after correction to be $\simeq 4.7\times10^{60}\rm\ erg$.

For a given outburst energy, the explosive shock model maximizes the
shock speed at all times, minimizing the estimated outburst age. A
model with constant jet power would produce slower shocks at early
times, although the shock speed would still decrease with time.
Assuming that the shock speed is constant at its present value
provides the likely upper limits on the outburst age of $2.34 \times
10^{7}\rm\ yr$ and $2.51 \times 10^{7}\rm\ yr$ for the northern and
southern shocks, respectively. Therefore, the age estimates from the
explosive shock model are unlikely to be low by more than a factor of
$\simeq 1.3$. Combining the total energy estimate of
$4.7\times10^{60}\rm\ erg$ with the upper and lower age estimates
gives estimates for the time-averaged outburst power in the range
0.6 -- 0.8 $\times10^{46}\rm\ erg\ s^{-1}$, in broad agreement with
other estimates \citep{Carilli1996, Wilson2006, Godfrey2013}. 
We use the value of $10^{46}\rm\ erg\ s^{-1}$, with a systematic 
uncertainty of a factor of 2, as representative of the mean 
outburst power below. Although the jet power, lobe, and ICM pressures 
of Cyg~A are high compared to those of the more typical FRII galaxies 
in the sample of \citet{Ineson2017}, its dimensionless properties, 
such as the pressure ratios and the cocoon shock Mach numbers, are typical.

%%%%%%%%%%%%%%%%%%%%%%%%%%%%%%%%%%%

\subsection{Hotspot Speeds}
\label{sec:hotspotspeed}

We can make a geometric estimate for the speed of advance of each 
hotspot. If the axis of the cocoon is inclined at $55^\circ$ to our
line of sight, the deprojected distance from the AGN to the shock near 
the tip of the eastern (western) jet is a factor of $\simeq 2.1$
($\simeq 2.5$) greater than the average distance from the  
AGN to the shock fronts in regions 1 and 5 (Table \ref{table:sbp}).
Multiplying these factors by the average speed of the advance for the
innermost shocks provides estimates of the time-averaged speeds of the
outermost parts of the cocoon shock.
Using an average shock speed in regions 1 and 5 of 
$1590 \pm 50 \rm\ km s^{-1}$, the recession speed of the 
eastern hotspot is $3340 \pm 110\rm\ km\ s^{-1}$ and that of
the western hotspot is $3980 \pm 130\rm\ km\ s^{-1}$. Using the
deprojected temperature of $6.42 \pm 0.27 \rm\ keV$ gives a
Mach number of $2.54 \pm 0.14$ for the shock near the eastern hotspot. 
With a deprojected temperature of $9.25 \pm 0.49 \rm\ keV$, the 
Mach number near the western hotspot is $2.52 \pm 0.16$. 

The estimated shock speeds fall well short of the hotspot speeds. 
In part, the shock speeds may be underestimated (Section~\ref{sec:uniformity}), 
but the shape of the shock front is also critical. 
We may use a self-similar model to demonstrate that this behavior is
consistent with a shock geometry that tapers toward the hotspots. 
We assume that the shape of the cocoon shock remains
fixed as it expands. Although this is an idealization, the
changes in relative speed that cause departures from self-similarity
generally occur on timescales comparable to the age of an outburst, so
we should not expect to find large departures from self-similar expansion in
practice. If the size of the front is proportional to $g(t)$, where
$t$ is the time, its shape projected onto the sky can be defined as a
level surface of a function of two arguments, in the form $f [x /
g(t), y / g(t)] = 0$, where the AGN is located at $x = y = 0$. Since
the shock velocity is perpendicular to the level surfaces of $f$, the
speed of the shock at any position on the front is given by
\begin{equation} \label{eqn:level}
v = \frac{1}{g} \frac{dg}{dt} \frac{|\mathbf{r} \cdot \nabla f|}
  {|\nabla f|} = \frac{v_0}{r_0} r \cos\theta,
\end{equation} 
where $\theta$ is the angle between the radius vector and the normal
to the front, so that $\cos\theta = \mathbf{r} \cdot \nabla f / (r
|\nabla f|)$. In the second form, values are referred to the position
on the front closest to the AGN, where $r=r_0(t)$, $v = v_0(t)$, and the
radius vector $\mathbf{r}$ must be perpendicular to the front, so that
$\cos\theta = 1$. This expression determines how the shock speed
depends on position at a fixed time. Alternatively, the shape of
the front may be specified by giving $x$ and $y$ as functions of a
parameter $s$ and then, from Equation~\ref{eqn:level}, the expansion 
speed varies over the front as
\begin{equation} \label{eqn:param}
v \propto r \cos\theta = \frac{|y dx/ds - x dy/ds|}{\sqrt{(dx/ds)^2 +
    (dy/ds)^2}}.
\end{equation}

Clearly, the expansion speed is the same at every point on a spherical front
(circular on the sky). It is also constant for a conical front of the form 
$y = y_0 - \eta x$, with constant $\eta$ (Equation~\ref{eqn:param}).
From Figure~\ref{fig:cyga}, the cocoon shock of Cyg~A may be roughly
approximated as a sphere, capped to the east and west by a pair of
opposed cones. If the cones are tangent where they attach to the
sphere, the expansion speed of the self-similar front would be
constant everywhere but at the tips of the two cones. Although this is a
crude model for the cocoon shock of Cyg~A, it illustrates how the
shock speed can be substantially less than the speed of the hotspots,
except in small regions close to the hotspots.

%%%%%%%%%%%%%%%%%%%%%%%%%%%%%%%%%%%

\subsection{Hotspot Pressures}
\label{sec:hotspotpres}

Rather than drilling into the ICM at a single location at the tip of
the cocoon, hotspots shift around rapidly \citep{Scheuer1982, Williams1985}, 
so that we expect the mean 
speed of the shock at the tip to be significantly lower than the
instantaneous speed of the hotspot. Therefore, using the mean speed
of the shock at the tip of the cocoon to estimate the hotspot pressure
should provide a minimum estimate, $\phsmin$. Using the value of the
external pressure at its deprojected distance of 77.0 kpc from the AGN
with the hotspot Mach number (Section \ref{sec:hotspotspeed}),
the pressure required to drive the eastern
hotspot needs to be at least $\phsmine = 1.48 \pm 0.32
\times10^{-9}\rm\ erg\ cm^{-3}$. For the western hotspot, at a
deprojected distance of 90.7 kpc, this calculation gives $\phsminw =
1.28 \pm 0.33 \times10^{-9}\rm\ erg\ cm^{-3}$.

Synchrotron self-Compton (SSC) models for the radio and X-ray emission
of the hotspots can provide more realistic estimates of the hotspot
pressures \citep{Harris1994}. Decomposing the pressure into a sum of
contributions from the magnetic field, electrons, and nonradiating
particles, it may be expressed as $p_{\rm hs} = p_{B} + p_{\rm e} +
p_{\rm n}$. To relate this to the results of the SSC model, we recast
it as
\begin{equation} 
  \label{eqn:p_hs}
  \phsp = \frac{U_{B}}{3} \left[\frac{3p_{B}}{U_{B}} +
    \frac{U_{\rm e}}{U_{B}} \left(1 + \frac{p_{\rm n}}{p_{\rm e}}\right)\right],
\end{equation}   
where $U_B$ is the magnetic energy density, $U_{\rm e}$ is the
electron energy density, and we have assumed that $p_{\rm e} = U_{\rm
  e}/3$ (tending to underestimate the electron pressure for
$\Gamma_{\rm min} \to 1$). Under the simplest assumptions, the
magnetic field is isotropic, so that the magnetic pressure is related
to the magnetic energy density by $p_B = U_B/3$, giving
\begin{equation} 
  \label{eqn:p_hs2}
  \phsp = \frac{U_{B}}{3} \left[1 + \frac{U_{e}}{U_{B}}\left(1 +
      \frac{p_{\rm n}}{p_{\rm e}}\right)\right]. 
\end{equation}  
For a light jet, electrons and positrons contribute equally to the
``electron'' pressure, $p_{\rm e}$, while the pressure of the
nonradiating particles is negligible. For a matter-dominated jet, the
number density of nonradiating particles (ions) in the hotspot will be
comparable to the electron density. Their relative pressures then
depend on the particle energy distributions, which are determined by
acceleration mechanisms. The simplest assumption would be
$p_{n}/p_{e} = 1$, but the acceleration mechanisms can also make the
ion pressure substantially greater than the electron pressure
\citep{Malkov2001}, so that the total hotspot pressure may be
substantially greater than our estimates below. If
the magnetic field is well-ordered, the effective magnetic pressure
could also be up to a factor of 3 greater. Higher hotspot pressures
would entail greater instantaneous hotspot speeds and larger mass
fluxes through the jets (see below and Section \ref{sec:jets}).

From their SSC model for the radio and X-ray emission of the bright
eastern hotspot (D), \citet{Stawarz2007} found $B = 270\rm\ \mu G$,
with values of $U_{\rm e}/U_B$ in the range 3 -- 4. Adopting $U_{\rm
  e}/U_B = 3.5$ as representative, Equation~\ref{eqn:p_hs2} gives
pressures of $\phse = 4.4 \times10^{-9}\rm\ erg\ cm^{-3}$ for a light
jet, or $\phse = 7.7 \times10^{-9}\rm\ erg\ cm^{-3}$ for a matter-dominated 
jet with $p_{\rm n} = p_{\rm e}$. Similarly, using $U_{\rm
  e}/U_B = 7.5$ and $B = 170\rm\ \mu G$ for the western hotspot (A)
gives corresponding pressures of $\phsw = 3.3 \times10^{-9}\rm\ erg\
cm^{-3}$ or $6.1 \times10^{-9}\rm\ erg\ cm^{-3}$. Bearing in mind the
substantial systematic uncertainties, these values are consistent with
expectations. For the two SSC pressure estimates above, the eastern
hotspot would drive shocks at Mach $\simeq 4.3$ or 5.7 into the ICM,
both significantly faster than the estimated mean Mach number of 2.54,
as anticipated. For the western hotspot, the SSC pressures would drive
shocks at Mach $\simeq 4.0$ or $5.4$, also both substantially faster
than the mean Mach number of 2.52.

These arguments relate values that may vary on widely differing
timescales. Jet fluxes are observed to change on timescales ranging
upward from the light-crossing time of the jet \citep[\eg,][]{Harris2006}. 
The hotspots of Cyg~A are compact, with radii of
$\simeq1$ kpc, and they are composed of gas that is likely to be
relativistic (or nearly so), so they can respond to rapid changes in
the confining pressure on timescales ranging
upward from a few thousand years. If the jet axis of Cyg~A is
inclined at $55^\circ$ to our line of sight, its western hotspots are
$\simeq 100$ kpc closer to us than its eastern hotspots, so that the
light travel time from the western hotspots is $\simeq 0.3$ Myr longer
(the delay may be partly offset by the greater distance from the
AGN to the western hotspots, but only orders of magnitude matter
here). The eastern jet of Cyg~A has a filamentary appearance,
with what appear to be twists over scales of several kiloparsecs
\citep{Perley1984}, suggesting it moves about on a timescale of several
thousand years (if the jet is relativistic). Thus, variations in jet
power or direction could cause the hotspot pressures 
determined from the SSC model to vary on timescales more than an order
of magnitude shorter than the light travel delay between the eastern
and western hotspots. By contrast, we have used the shock speed near
the tips of jets averaged over the duration of the outburst, about 20
Myr (Section~\ref{sec:ageenergy}), for the other pressure estimates.
Given the disparity of the timescales, the consistency between the
various pressure estimates is, perhaps, fortuitous. It implies that the
``current'' values of the jet power at both hotspots are comparable to
the mean power averaged over the lifetime of the outburst.

%%%%%%%%%%%%%%%%%%%%%%%%%%%%%%%%%%%

\subsection{Jet and Hotspot Composition}
\label{sec:jets}

\begin{table*}
	\caption{Sample Jet Model Parameters}
	\label{tab:models}
	\begin{tightcenter}
	\begin{tabular}{c c c c c c c c c c c}
		\hline \hline 
		Jet & $\pjet$\tablenote{Jet power is reduced below 
			$5 \times 10^{45} \rm\ erg\ s^{-1}$ in cases where 
			the minimum thrust exceeds the thrust estimated from the 
			hotspot pressure. The resulting maximum value 
			makes $\dot M = 0$.} 
			& $\Gamma$ & $p$ 
			& $\phsp$ & $\rjet$\tablenote{Jet area is $A = \pi \rjet^2$.}
			& $\beta$ & $\dot M$ & $f_{\rm KE}$	\tablenote{Kinetic 
			power fraction is $(\gamma - 1) \dot M c^2 / \pjet$.} 
			& $\kTjet$	\tablenote{Equivalent jet temperature for 
			matter-dominated models, 
			$\kTjet = \mu m_{\rm H} p / \rho$, with $\rho$ from
			Equation~\ref{eqn:mdot}.}  & $s/c$\tablenote{Sound 
			speed from the Synge model for hydrogen plasma
			at temperature $\kTjet$ for matter-dominated models and 
			$c /\sqrt{3}$ for light jet models.}   \\
  		& ($10^{45}\rm\ erg\ s^{-1}$) & & 
			\multicolumn{2}{c}{($10^{-9}\rm\ erg\ cm^{-3}$)} & (kpc) 
			& & ($\rm M_\odot\ yr^{-1}$) & & (MeV) & \\ 
		\hline
		East & 5 & 13/9 & 1.04 & 7.7 & 1.0 & 0.667 & 0.070 & 0.27 & 118 & 0.438 \\
		East & 5 & 13/9 & 1.04 & 7.7 & 0.5 & 0.882 & 0.031 & 0.40 & 137 & 0.455 \\
		East & 4.51 & 4/3 & 1.04 & 4.4 & 1.0 & 0.668 & 0 & 0 & --- & 0.577 \\
		East & 3.45 & 4/3 & 1.04 & 4.4 & 0.5 & 0.874 & 0 & 0 & --- & 0.577 \\
		West & 5 & 13/9 & 0.84 & 6.1 & 1.2 & 0.608 & 0.109 & 0.32 & 75 & 0.385 \\
		West & 5 & 13/9 & 0.84 & 6.1 & 0.6 & 0.835 & 0.055 & 0.51 & 73 & 0.382 \\
		West & 4.89 & 4/3 & 0.84 & 3.3 & 1.2 & 0.650 & 0 & 0 & --- & 0.577 \\
		West & 3.68 & 4/3 & 0.84 & 3.3 & 0.6 & 0.863 & 0 & 0 & --- & 0.577 \\
		\hline
	\end{tabular}
	\end{tightcenter}
\end{table*}

The power and pressure estimates can be used to estimate some
properties of the jets. Given the substantial systematic
uncertainty, we employ a one-dimensional, steady, relativistic
flow model \citep{Landau1959, Laing2002}. The flow rate of rest mass
through the jet is given by
\begin{equation} 
  \label{eqn:mdot}
  \dot{M} = \rho A c \beta \gamma,
\end{equation} 
where $\rho$ is the proper density of the rest mass, $A$ is the
cross-sectional area of the jet, the bulk flow speed is $v = \beta c$,
and $\gamma$ is the corresponding Lorentz factor. The jet power can be
expressed as
\begin{equation} 
  \label{eqn:jetpower}
  \pjet = (\gamma - 1)\dot{M}c^2 + h A c \beta \gamma^2,
\end{equation}
where the enthalpy per unit volume is related to the pressure by $h =
\Gamma p / (\Gamma -1)$ and the ratio of specific heats
of the jet fluid, $\Gamma$, is assumed to be constant. The total
momentum flux, or thrust, of the jet is given by 
\begin{equation} 
  \label{eqn:jetmoment}
  \Pi = (\pjet / c + \dot{M} c) \beta.
\end{equation}
For both hotspot A and hotspot D, the SSC pressure estimate of
Section~\ref{sec:hotspotpres}  significantly exceeds our estimate of the
static pressure in the lobe. 
Assuming that the excess pressure in a hotspot is due to the ram
pressure of the jet, we have 
\begin{equation}
  \label{eqn:thrust}
  \Pi = (\phsp - p) A_{hs},
\end{equation}
where $A_{\rm hs}$ is the hotspot cross-sectional area. Although neither 
A nor D is a ``primary'' hotspot \citep{Carilli1996}, each is the largest and 
brightest in its lobe and so provides the greatest estimates for the jet thrust.
This estimate would decrease under adiabatic expansion, so that, if
the hotspot is no longer confined by the jet, our estimate of the jet
thrust will be low, causing the mass flow rates to be underestimated.
Using Equation~\ref{eqn:jetpower} to eliminate $\dot M$ in
Equation~\ref{eqn:jetmoment} then yields
\begin{equation}
  \label{eqn:beta}
  \frac{\pjet}{\Pi c} = \frac{\beta \gamma}{\gamma + 1} +
  \frac{hA}{\Pi} \beta\gamma
\end{equation} 
and this can be solved for the flow speed using the estimates above.

For each lobe, we use the SSC estimates of the hotspot pressure for
$p_{\rm n} = 0$ and for $p_{\rm n} = p_{\rm e}$
(Section~\ref{sec:hotspotpres}). For the static pressure in each jet, we
use the lobe pressure estimated from the averaged rim pressure
(Section~\ref{sec:rimpressure}). The jet power was taken to be one-half 
of the mean total outburst power, so we use $\pjet =
5\times10^{45}\rm\ erg\ s^{-1}$ as a representative value
(Section~\ref{sec:ageenergy}). However, the minimum jet thrust
would then exceed the thrust estimated from the hotspot pressure 
for the lower hotspot pressure in each lobe (for $p_{\rm n} = 0$). For
those cases, the jet power was reduced to the maximum
value consistent with the hotspot pressure,
\begin{equation}
  \label{eqn:pjmax}
  \pjmax = c \sqrt{(\Pi + hA) \Pi},
\end{equation} 
which makes $\dot M = 0$.

The higher hotspot pressure estimates were obtained under the
assumption that the nonradiating particles contribute as much to the
hotspot pressure as the electrons, in which case we should expect the
hotspot, hence also the jet, to be matter dominated. For the model
parameters discussed here, the equivalent temperatures of the jet
($\kTjet = \mu m_{\rm H} p / \rho$) fall in the MeV to GeV range,
so the electrons would be relativistic and most of the ions
nonrelativistic, making the ratio of specific heats for the jet close
to $\Gamma = 13/9$. By contrast, to obtain the lower hotspot
pressures, the pressure of the nonradiating particles in the hotspot
is assumed to be negligible. This implies the jet has a negligible
ion content, with positrons as the predominant positive charges. Such
a jet is light, and the majority of particles in it will be relativistic,
making $\Gamma = 4/3$  \citep{Krause2003, Krause2005, Guo2011}.
\cite{Kino2012} found both heavy and 
light jet models to be consistent with prior Cyg~A observations, 
albeit with a preference for light jets, and so we considered 
both models in our analysis.

Since the jet widths are difficult to assess from the 5 GHz radio map,
we relate them to the hotspot sizes. In both the X-ray and 5 GHz
radio images, we estimate the FWHMs of hotspots
A and D to be 2.4 kpc and 2.0 kpc, respectively. If the jet covers
the whole of each hotspot, the corresponding jet radii would be about
1.2 kpc in the west and 1.0 kpc in the east. A jet may also be
narrower than its hotspot, confining it by the dentist drill effect
\citep{Scheuer1982}. To keep it confined, the moving tip of the jet
must then traverse the whole hotspot within the few thousand years
required for the hotspot to expand significantly. From
Section~\ref{sec:hotspotpres}, the twisted appearance of the eastern jet
suggests it moves on a timescale of several thousand years. To keep
the hotspots confined, we therefore assume that the jet radius needs
to be at least half that of the hotspot, and so we use jet radii equal to
0.5 or 1 times the radius of the hotspot.

The two jet radii and two hotspot pressures for each lobe, with their
accompanying jet powers and equations of state discussed above, give
the eight sets of model parameters and results listed in
Table~\ref{tab:models}. The four matter-dominated heavy jet models
give jet speeds in the range $\beta = 0.61$ -- 0.88, while those for
the light jet models are in the range $\beta = 0.65$ -- 0.87. These
are comparable to VLBI speeds measured in the core
\citep{Krichbaum1998, Boccardi2016}, although the large uncertainties
in the powers and areas of the jets would allow almost any speed
$\gtrsim 0.15 c$ (jet speeds with large $\gamma$ requiring much
smaller jet areas). For the light jet models, the jet powers, reduced
to be consistent with the hotspot pressures
(Equation~\ref{eqn:pjmax}), lie in the range \mbox{3.4 --
$4.9\times10^{45}\rm\ erg\ s^{-1}$}, comfortably in agreement with the
results of Section~\ref{sec:ageenergy}. Our results do not rule out
light, matter-dominated jets (with a higher hotspot pressure, $\pjet
\simeq \pjmax$ and $\Gamma = 4/3$), although they would require powers of
\mbox{6.5 -- $8.7\times10^{45}\rm\ erg\ s^{-1}$}, stretching the upper limit
of the acceptable power range at the higher end. Jet speed is an 
increasing function of the power and a decreasing function of the
thrust and jet area. For the matter-dominated models, the sound
speeds in the jet ranges from \mbox{0.38 -- $0.45 c$} and the jet Mach
numbers from \mbox{1.5 -- 2.2}. For the light jet models in the table, the
sound speed is $c / \sqrt{3}$ and the range of Mach numbers is 
\mbox{1.13 -- 1.50}.

Considering only models with jet speeds comparable 
to the VLBI speeds, the kinetic power fraction,
$(\gamma - 1) \dot M c^2 / \pjet$, is modest. For example, the 
matter-dominated models of Table~\ref{tab:models} have kinetic power
fractions $\lesssim 50\%$. As defined here, the kinetic power
fraction of the light jet models is zero, since $\dot M = 0$.
Restricting attention to these models eliminates the poorly
constrained parameter $\dot M$, while still yielding properties
consistent with our observations. The jet power for the light jet models
is $\pjmax$ (Equation~\ref{eqn:pjmax}), which is determined by the
other model parameters. It is most sensitive to our estimate of the
hotspot pressure and somewhat less sensitive to the jet pressure and
area. The resulting jet powers are consistent with the estimates from
Section~\ref{sec:ageenergy} and the corresponding jet speeds,
\begin{equation}
\beta = \sqrt{\frac{\Pi}{\Pi + h A}},
\end{equation}
are also in the range expected.

The jets of Cyg~A may well have entrained some ordinary matter.
For example, assuming they are old, the stars of Cyg~A would shed
roughly $5\times 10^{-5}\rm\ M_\odot\ yr^{-1}$ within the volume of
each jet. If all of this is entrained, it would contribute $\dot M
c^2 \simeq 3 \times 10^{42}\rm\ erg\ s^{-1}$ to each jet, three orders
of magnitude smaller than the jet power. From
Equations~\ref{eqn:jetpower} and \ref{eqn:jetmoment}, this amount of
entrained gas would only have an appreciable impact on the flow for
$\gamma \gtrsim 1000$. For $\gamma \sim 1$, unless the mass entrained
by the jets is about three orders of magnitude greater, the jets may
be treated as light. Since light jet models provide flow solutions
that are consistent with the observed properties of Cyg~A, these
results favor the jets being light.

%%%%%%%%%%%%%%%%%%%%%%%%%%%%%%%%%%%

\section{Conclusions}
\label{sec:conclusion}

Deep \chandra{} observations of the cocoon shock of Cyg~A were analyzed
to quantify physical properties of the AGN outburst, the
lobes, and the jets of Cyg~A. X-ray surface brightness profiles of the
shocks were used to determine shock strengths in a number of regions
around the cocoon. Fitting the profiles with a hydrodynamic model for
the AGN outburst gave Mach numbers for the cocoon shock in the range
1.18--1.66. The outburst energy for the system was determined to be
$\simeq 4.7\times10^{60}\rm\ erg$, after substantial corrections, and
the outburst age was found to be $\simeq 2 \times 10^{7}\rm\ yr$,
giving a mean outburst power of $P \simeq 10^{46}\rm\ erg\ s^{-1}$,
with a systematic uncertainty of about a factor of 2. The mean power
is consistent with independent estimates of the outburst power for
Cyg~A based on simulations of radio and X-ray emissions.

The off-center location of the AGN with respect to the cocoon shock
indicates that it is moving through the ICM. From the shock speeds
and age, the AGN (i.e., the BCG) is estimated to be moving at $270\rm\ 
km\ s^{-1}$ with respect to the gas, with a substantial systematic 
uncertainty. 

Spectra of regions in the thin rim of compressed gas between the radio
lobes and the shocks were used to estimate pressures. The mean rim
pressure agrees well with the postshock pressures determined from the
shock jump conditions. The rim pressure for the western lobe, $8.4\pm
0.2\times10^{-10}\rm\ erg\ cm^{-3}$, is $\sim20\%$ lower than the mean
value for the remainder of the cocoon, $10.4\pm
0.4\times10^{-10}\rm\ erg\ cm^{-3}$. The rim pressures provide good
estimates of the pressure within the radio lobes, apart from the
vicinity of the hotspots. They show some evidence for persistence of a
20\% pressure difference between the east and the west, which is
puzzling given the high sound speed expected in the lobes. Despite
this, one of our main findings is that the pressure is
uniform within $\sim$20\% throughout the bulk of the cocoon.

Scaling by distance from the AGN, we estimate the Mach numbers of the
shocks near the hotspots of Cyg~A
to be $2.54\pm0.14$ in the east and $2.52\pm0.16$ in the west,
significantly greater than any Mach number obtained by fitting the cocoon
shock. A simple geometric model shows that
the shock speed need only be so high in a small region close to the
hotspots. The speed of the shock front near the hotspots
places lower limits on the hotspot pressures of $\phsmine = 1.48 \pm
0.32 \times10^{-9}\rm\ erg\ cm^{-3}$ in the east and $\phsminw = 1.28
\pm 0.33 \times10^{-9}\rm\ erg\ cm^{-3}$ in the west. These values
are higher than the estimated cocoon pressure but are significantly lower
than the hotspot pressures estimated from SSC
models. This is consistent with positions of the hotspots moving
about on the cocoon shock over time. 
The SSC-derived hotspot pressures show that the ram pressures of the
jets are at least twice as large as their 
static pressures.

Estimates of the jet power and hotspot pressures were used with a
steady, one-dimensional, matter-dominated flow model to determine jet
properties. These models are consistent with mildly relativistic flow 
speeds within the allowed parameter ranges. Notably, light jet
models, which carry a negligible flux of rest mass and so have one
less parameter than the general model, agree with the
observed properties of the jets and hotspots. This result favors the jets of 
Cyg A being light, meaning that both the momentum flux and kinetic 
power due to the flow of rest mass through the jets are negligible 
compared to those due to the flow of internal energy.

\acknowledgments{
Support for this work was provided by the National Aeronautics and Space
Administration through Chandra Award Number GO5-16117A issued by the 
Chandra X-ray Observatory Center, which is operated by the Smithsonian 
Astrophysical Observatory for and on behalf of the National Aeronautics 
Space Administration under contract NAS8-03060.
P.E.J.N. was supported in part by NASA contract NAS8-03060.}

%%%%%%%%%%%%%%%%%%%%%%%%%%%%%%%%%%%

\bibliographystyle{aasjournal}
\bibliography{all_data}

\end{document}